\documentclass[%
groupedaddress,
 amsmath,amssymb,
 aps,
prb,
twocolumn,
floatfix,
notitlepage
]{revtex4-1}

\usepackage{graphicx,placeins}
\usepackage{dcolumn}
\usepackage{bm}
\usepackage{subfigure}
\usepackage{units}
\usepackage[usenames, dvipsnames]{color}
\usepackage[colorlinks=true,bookmarks=true,%
     pdfborder={0 0 0},
     breaklinks=true,bookmarksnumbered=true]{hyperref}
     \hypersetup{
    colorlinks,
    citecolor=blue,
    filecolor=cyan,
    linkcolor=blue,
    urlcolor=magenta
}

\usepackage{atveryend}
\makeatletter
\let\origcitation\citation
\AtEndDocument{\def\mycites{}%
  \def\citation#1{\g@addto@macro\mycites{#1^^J}\origcitation{#1}}}
\AtVeryEndDocument{\newwrite\citeout\immediate\openout\citeout=\jobname.cit
  \immediate\write\citeout{\mycites}\immediate\closeout\citeout}
\makeatother

\def \pt{\partial}

\def \beq{\begin{equation}}
\def \eeq{\end{equation}}
\def \beqarr{\begin{eqnarray}}
\def \eeqarr{\end{eqnarray}}
\def \bspt{\begin{split}}
\def \espt{\end{split}}
\def \bef{\begin{figure}}
\def \enf{\end{figure}}

\def \bpm{\begin{pmatrix}}
\def \epm{\end{pmatrix}}

\newcommand {\apgt} {\ {\raise-.5ex\hbox{$\buildrel>\over\sim$}}\ }
\newcommand {\aplt} {\ {\raise-.5ex\hbox{$\buildrel<\over\sim$}}\ }
\newcommand{\lessim}{\aplt}

\newcommand{\si}{\sigma}

\newcommand{\tJ}{\ $t$-$J$ \ }
\newcommand{\barray}{\begin{eqnarray}}
\newcommand{\earray}{\end{eqnarray}}
\newcommand{\nn}{\nonumber}
\newcommand{\disp}[1]{Eq.~(\ref{#1})}

\newcommand{\refdisp}[1]{Ref.~[\onlinecite{#1}]}
\newcommand{\figdisp}[1]{Fig.~\ref{#1}}
\renewcommand{\O}{{\cal O}}

\newcommand{\abs}[1]{\lvert#1\rvert}

 \definecolor{indigo}{rgb}{0.5,0.1,1}

\begin{document}


\title{ A Strange Metal  from Gutzwiller correlations  in infinite dimensions}

\author{ Wenxin Ding$^1$, Rok \v{Z}itko $^{2,3}$, Peizhi Mai$^1$, Edward Perepelitsky$^1$ and B Sriram Shastry$^1$}

\affiliation{
$^1$Physics  Department, University of California, Santa Cruz, California, 95060,\\ $^2$
Jo\v{z}ef Stefan Institute, Jamova 39, SI-1000 Ljubljana, Slovenia\\$^3$Faculty for Mathematics and Physics, University of
Ljubljana, Jadranska 19, SI-1000 Ljubljana, Slovenia
 }%

\date{\today}
\begin{abstract}
Recent progress in {\em extremely correlated Fermi liquid theory
(ECFL)} and the {\em dynamical mean field theory (DMFT)} enables us to
accurately compute  in the $d \rightarrow \infty$ limit the
resistivity of the $t$-$J$ model after setting $J\to0$. This is also  the  $U= \infty$ Hubbard
model. Since $J$ is set to zero,  our study isolates the {\em dynamical effects} of the
single occupation constraint enforced by the projection operator originally introduced by Gutzwiller. We study three densities
$n=.75,.8,.85$ that correspond to a range between the overdoped and
optimally doped Mott insulating state.    We delineate four distinct
regimes separated by three crossovers, which are characterized by
different behaviors of the resistivity $\rho$. We find at the lowest
temperature $T$ a  {\em Gutzwiller correlated Fermi liquid} regime with
$\rho \propto T^2$ extending up to an effective Fermi temperature that
is dramatically suppressed from the noninteracting value by the
proximity to half filling, $n\sim 1$.  This is followed by a  {\em Gutzwiller correlated strange metal} regime with  $\rho \propto (T-T_0)$, i.e., a linear resistivity extrapolating back to $\rho=0$  at a positive  $T_0$. At a higher temperature scale  this crosses over into the {\em bad metal} regime with $\rho \propto (T+T_1)$, i.e., a linear resistivity extrapolating  back to  a finite resistivity at $T=0$ and passing through the Ioffe-Regel-Mott value where the mean free path is a few lattice constants. This regime finally gives way to the {\em high $T$ metal} regime, where we find $\rho \propto T$, i.e., a linear resistivity extrapolating back to zero at $T=0$.
The present work emphasizes the first two, i.e. the two lowest temperature  regimes, where the availability of an analytical ECFL theory is of help in identifying the changes in related variables entering the resistivity formula  that accompanies the onset of linear resistivity, and the numerically exact DMFT helps to validate the results.  We also examine thermodynamical variables such as the magnetic susceptibility, compressibility, heat capacity, and entropy and correlate changes in these with the change in resistivity.  This exercise  casts valuable light on  the  nature of charge and spin correlations in the  Gutzwiller correlated strange metal regime, which has features in common with the physically relevant strange metal phase seen in strongly correlated matter.

\end{abstract}

\pacs{Valid PACS appear here}
\maketitle



\section{Introduction}

The resistivity due to mutual collisions of electrons at low
temperatures reveals the lowest energy scale physics of charge
excitations in metallic systems, and therefore is very important.
While it is fairly straightforward to measure experimentally, it is
also one of the most difficult quantities to calculate theoretically,
especially if electron-electron interactions are strong. Motivated by
the unexpected behavior of resistivity and other variables in cuprate
superconductors and related two-dimensional experimental systems, some
works have postulated that the Fermi liquid theory - originally
developed and justified for weakly interacting systems - would break
down. In its place a zoo of non-Fermi liquids have been postulated,
without necessarily having a rigorous theoretical underpinning. On the other hand the analytical
framework of the extremely correlated Fermi liquid theory (ECFL)
\cite{ECFL} and the well established dynamical mean field theory
(DMFT) \cite{DMFT} give a different type of result, where the strong
interactions compress the regime of Fermi-liquid type variation to a
very small temperature and frequency scale. This Fermi-liquid regime
is succeeded by a variety of regimes that display unusual
non-Fermi-liquid dependences on frequency and temperature. The main
goal of this work is to elucidate and characterize the different
regimes that arise in the ECFL and DMFT theories, and to provide a
quantitative comparison between the qualitatively similar results of
these two theories, as applied to the infinite-dimensional Hubbard
model, with the Hubbard charge repulsion parameter $U$ taken to
infinity, $U\to \infty$.

In earlier work\cite{ECFL-DMFT} we have compared the ECFL and DMFT
results for the zero-temperature spectral functions, finding an
encouraging similarity. On scaling the frequency with the respective
quasiparticle weights $Z$ of the two theories the agreement is even
close to quantitative.  In the present work we undertake the more
ambitious comparison of the resistivity and thermodynamic variables at
finite temperatures.

In both the ECFL theory and the DMFT, the strong interactions cause 
the quasiparticles of the lowest temperature Fermi liquid to become
fragile, i.e., the resulting quasiparticle weight $Z$ is very small,
$Z \ll 1$. This is also arguably the relevant regime in contemporary
materials such as cuprate superconductors, and hence interest in this
problem is very high.

In the problem studied here, namely $U\to\infty$ and $d\to\infty$, the
DMFT theory is formally exact. Further, the possibility of computing
the resistivity from the sole knowledge of the single-particle Green's
function is enabled by the vanishing of vertex
corrections\cite{Khurana}.  Despite these simplifications, obtaining
{\em reliable} results for the resistivity is technically formidable
due to the requirement of an impurity solver providing accurate and
reliable results for the self-energy function $\Sigma$ on the real
frequency axis for both very low and very high temperatures. This
problem has only recently been solved in \refdisp{badmetal}, almost 25
years after the formulation of the DMFT theory. The resistivity of the
Hubbard model is now known for all densities and all values of U,
including $U=\infty$. This is a set of exact results for the
resistivity in interacting metallic systems resulting from inelastic
scattering, and therefore represent an important advance in the field.
The DMFT results \cite{badmetal,HFL} offer a unique opportunity to
test a variety of techniques and approximate methods for computing
this variable. The ECFL formalism, on the other hand, is in its early
stages of development and several technical innovations are ongoing so
as to enable reliable calculations in the challenging regimes of the
density $n \lessim 1$ \cite{ECFL-DMFT,ECFL-resistivity}.

Lastly, in a  recent work \refdisp{high-T-resistivity}  our group has  published  a voluminous
high-temperature study   using series expansion techniques  adapted for
very strong correlations,
 thus  extending  our understanding of the
resistivity to the full range of temperatures.
This study is on the same model as the present work and
 extends the results of \refdisp{badmetal} to much higher
temperatures. In these studies the effect the superexchange $J$
is absent due to the $U=\infty$ limit,  and therefore there is no
superconducting  regime that one might expect from a \tJ model in finite dimensions.
By taking the limit of infinite $U$ we have also banished the static superexchange that the DMFT includes for finite $U$ \cite{mullerhartmann1989,jarrell1992dmft,strack1992,Jarrell:1994ut,fleck1998,zitzler2002,pruschke2003,sangiovanni2006static,peters2007magnetic}.
However, these
studies {\em do} capture the notoriously difficult nonperturbative
local Gutzwiller correlation effects on the resistivity
quantitatively.  It seems fair to say that our
understanding of the strong correlation problem has advanced
significantly with these recent works.

\begin{figure}[h]
  \centering
  \includegraphics[width=\columnwidth]{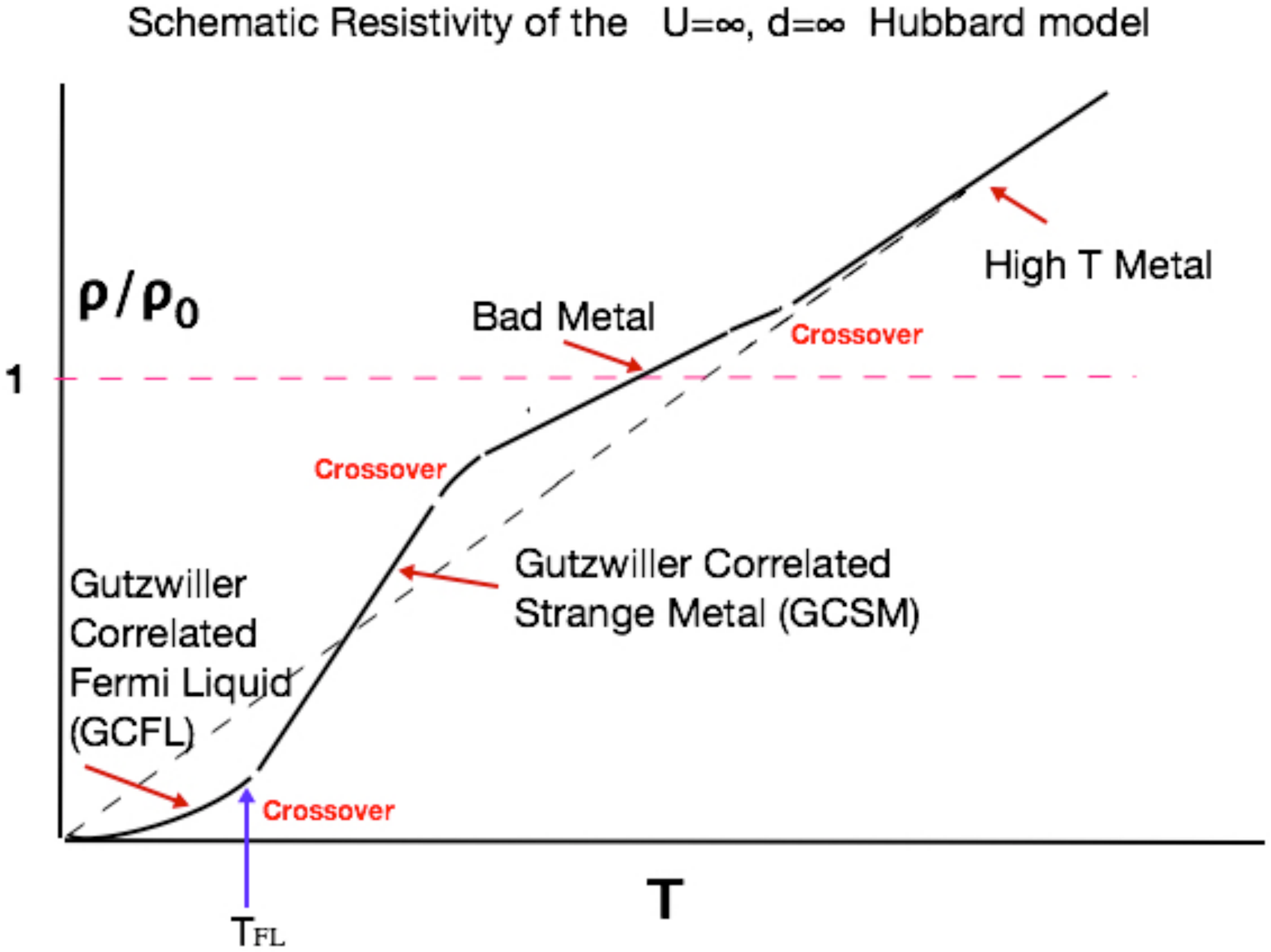}
  \caption{A schematic view of the different regimes of temperature
  dependent resistivities found in the calculations of
  \refdisp{badmetal,HFL,ECFL-resistivity,high-T-resistivity}.  The
  various temperature scales are schematic. At the lowest $T$ we have a
  Gutzwiller-correlated-Fermi liquid regime (GCFL) with $\rho \propto
  T^2$. This quadratic variation terminates at a characteristic Fermi
  temperature $T_{FL}(\delta)$, which is found to be surprisingly
  small relative to $T_{BR}= \delta D$, the Brinkman-Rice temperature
  scale ($2D$ is the bandwidth). Upon warming we reach the
  Gutzwiller-correlated-strange metal (GCSM) regime, which is the main
  focus of this work. This gives way at higher $T$ to the so-called
  bad-metal regime with a resistivity that increases linearly beyond
  the Ioffe-Regel-Mott value $\rho_0$ characteristic of disordered
  metals. The temperature scale of this regime is $T_{BR}$ discussed above.
  Finally at the highest $T$ we reach the high $T$ regime with $\rho
  \propto T$ that can be extrapolated back to pass through the origin.
  We thus find a total of four regimes separated by three crossovers.
  It should be noted that in both theories considered here, the
  approximate range of the temperatures scales are $T_{FL}\sim
  0.004-0.01 D$, and the crossover to the bad-metal regime occurs at
  $T\sim 0.04-0.06 D$ for the densities considered ($n=0.75$ to
  $n=0.85$). }
  \label{fig:schematic-rho}
\end{figure}

 In summary, at the lowest temperatures these earlier studies
 \cite{badmetal,HFL,ECFL-resistivity,high-T-resistivity} found a Fermi-liquid type resistivity with $\rho
\propto T^2$. This regime extends   only up to $T_{FL}(\delta)$, a Fermi-liquid temperature scale dependent on the hole density ($\delta \equiv1-n$).  We shall term this the {\em Gutzwiller correlated Fermi liquid} (GCFL) regime.
This regime is followed by three distinguishable regimes with linear in $T$ resistivity having different slopes and intercepts, which are separated by crossovers;  a {\em Gutzwiller correlated strange metal} (GCSM) followed by a ``bad metal'' and finally a ``high-T metal''
regime, as   discussed below (see \figdisp{fig:schematic-rho}).
 The nomenclature stresses  that  these regimes  originate purely from
 Gutzwiller correlations (i.e., double occupancy avoidance).  In
 particular the regimes have no dependence upon  the superexchange energy $J$ or other energy scales which might be  additionally involved  in producing  the related strange metal  found in cuprates \cite{phase-diagram-hightc,Gutzexplanation}.

In order to understand the low-temperature regimes,
  we  would like to throw light on the factors that lead to extraordinarily low values of the Fermi temperature $T_{FL}(\delta)$ that are found.
 We  also wish to provide a detailed  understanding of the behavior of constituent variables  that lead to a linear resistivity  in the GCSM regime, starting at this low temperature. Here the ECFL theory provides us with a great advantage since it is largely analytical, and one can inspect the various constituents in detail.
  It is also interesting to seek a possible causal relationship between the linear temperature dependence of $\rho$ in the GCSM regime and the nature of incipient order (either spin or charge) that might be present. For this purpose, it is useful to compute,   by using
the techniques of \refdisp{badmetal,ECFL-resistivity}, the entropy and heat capacity,   the magnetic susceptibilities and compressibility. For completeness we also study the thermoelectric transport, as well as a few dynamical quantities such
as the self energy of the electrons. In a following paper we present other dyamical variables such as the optical conductivity.
These quantities provide a
complete picture of the metallic states having various temperature dependences
sketched in \figdisp{fig:schematic-rho}.

The lowest temperature {\em Gutzwiller-correlated Fermi liquid} (GCFL)
with $\rho \propto T^2$ shows enhancements of certain static
susceptibilities that are similar to those of the normal state of
liquid $^{3}$He. The {\em almost localized Fermi liquid theory} (ALFL)
of these enhancements is discussed by Vollhardt, W{\"o}lfle, and
Anderson in \refdisp{Vollhardt-RMP,Vollhardt-Anderson} on the basis of
 Gutzwiller's wave function and its approximation to the Hubbard model, where the variation
of the Landau parameters with density at fixed (large) U is
considered.  In particular \refdisp{Vollhardt-Anderson} studies the
enhancements of Fermi liquid parameters leading to enhanced effective
mass $m^*/m$, magnetic susceptibility $\chi_{spin}/\chi^{0}_{spin}$
and the bulk modulus (i.e., the inverse compressibility). 
Within the ALFL all three stated enhancements {\em are} proportional to the
inverse of $Z$ in that theory as well as in $^{3}$He.  We  check
below the extent to which this is true in the GCFL regime, to see how
it compares with the predictions of the ALFL theory, and find that the behavior of the compressibility is somewhat different.

   Upon warming we reach the GCSM regime with a linear temperature
dependence of the resistivity $\rho$.  This regime is interesting
since it is reminiscent of the {\em strange metal} regime in the
cuprate phase diagrams \cite{phase-diagram-hightc}.  It is remarkable
that this linear resistivity regime extends to very low T, essentially
the $T_{FL}(\delta)$, and one wants to know if this behavior is
causally linked to a change in entropy, i.e. to disordering. We aim at
correlating the GCSM regime with the extent of short ranged spin or
charge order in this regime.  These should be reflected in the heat
capacity and the entropy gain. By computing these variables, we show
that upon warming from $T=0$ substantial entropy is released as we
reach $T_{FL}$. However in the entire GCSM regime the magnetic
susceptibility is Pauli like, i.e., with an approximately $T$
independent behavior, and hence spin entropy should be unchanged. From
a high-$T$ expansion and on various general grounds, it is known that
it changes into a Curie-Weiss type behavior at the onset of the
bad-metal regime. 


The GCSM regime is followed by other subtly different $T$ dependences
as described in Sec.~\ref{resistivity} which are obtained in the bad-metal
regime and the high-temperature regime.  The density dependences of
the various crossover scales give important insight into the physics
of the resistivity. With one exception, all calculations reported here
are performed using both ECFL and single-site DMFT methods. Using the
two methods is very important since it gives us the opportunity to
benchmark the mostly analytical and relatively new ECFL technique with
the established and largely numerical DMFT method. The magnetic
susceptibility is available only from DMFT, and our presentation below
seems to be the most extensive result for this subtle variable
reported to date \cite{Jarrell:1994ut,kajueter1996,bauer2009}.

The plan of the paper is as follows. In Sec.~\ref{sec-methods} we
first make some further technical remarks about the methods. In
Sec. \ref{resistivity} we describe the various $T$ dependences of the
resistivity which serve to define the GCFL and GCSM regimes, and also
point to the higher $T$ bad-metal and high-T regimes. In Sec.
\ref{compressibility} we compare the chemical potential and
compressibility.  In Sec. \ref{bubbles} we discuss the frequently
made bubble approximation for the charge and spin susceptibilities,
and show that the bubble susceptibility is exactly expressible as an
integral of the energy derivative of momentum distribution function in
$d=\infty$. We also note that it is a good approximation to the exact
result for the charge susceptibility, but not so for the spin
susceptibilty. In Sec. \ref{selfenergy} we illustrate the self
energy and local density of states from the two methods, and find
that within ECFL the quasiparticles tend to have somewhat smaller Z at
the highest densities, as compared to DMFT. This causes a few other
differences described later. In Sec. \ref{heatcapacity} we examine
further $T$ dependent properties, the heat capacity and entropy.
Sec. \ref{susceptibility} discusses the magnetic susceptibility
$\chi$ from the DMFT calculations and lists some of the technical
difficulties that prevent its evaluation in the ECFL theory.  In
Sec. \ref{thermoelectric} we discuss the thermoelectric transport
coefficients, the Seebeck coefficient and the Lorenz number as well as
the thermoelectric efficiency. In Sec. \ref{sec-2} we discuss the
salient features of our results.

\FloatBarrier

\section{Methods}
\label{sec-methods}
  
In ECFL we have thus far used an expansion in the parameter $\lambda$,
which plays a role analogous to the quantum parameter $\frac{1}{2 S}$
in quantum theories of magnetism, where $S$ is magnitude of the spin.
In the first DMFT-ECFL comparison paper \refdisp{ECFL-DMFT}, we used
the second order terms in an expansion in $\lambda$. This
approximation led to quantitatively reliable answer for the
quasiparticle weight $Z$ at low temperature only in the overdoped
regime $n \lessim .75$, but to a nonvanishing value of $Z$ for $n\to
1$. In the more recent paper \refdisp{ECFL-resistivity} this problem
was addressed using the exact, rather than the $\lambda^2$ version of
the hole number sum rule, together with a cut-off for the tails of the
spectral function at very high energies. This procedure extends the
validity of the second order terms to higher density $n \lessim 0.85$,
so that the $Z$ values at low $T$ tend to zero as the insulating state
is approached and are comparable to, if somewhat smaller than, the
DMFT results. Due to this improvement, we found that the resistivity
is now on the same scale, and exhibits very similar crossover features
as the results in \refdisp{badmetal,HFL}, as detailed below. In this
work we report the comparison between the $T$ dependent resistivity and
other thermodynamic variables found from this cutoff scheme
\cite{cutoff-comment} and the exact results from DMFT. {We use the Bethe lattice  semicircular density of states $D(\epsilon)= \frac{2}{\pi D} \sqrt{1-\epsilon^2/D^2}$  in both theories.  }

The ECFL scheme used here has been described in detail in
\refdisp{ECFL-resistivity}, and consists of using the $\O(\lambda^2)$
expansion with the full number sum rule and the Tukey window used to
cut off the spectral width at very high energies. 

The DMFT scheme has been described in detail in \refdisp{ECFL-DMFT}.
The NRG calculations \cite{wilson1975,bulla2008} in this work were
performed with the discretization parameter $\Lambda=2$, using the
discretization scheme from Refs.~\onlinecite{zitko1,zitko2} with
$N_z=16$ interleaved discretization grids. The truncation cut-off was
set at $10\omega_N$, where $\omega_N$ is the characteristic energy
scale at the $N$-th step of the iteration. We used charge conservation
and spin SU(2) symmetries. The spectral functions were computed with
the full-density-matrix algorithm \cite{weichselbaum} and broadened
with a log-Gaussian kernel with $\alpha=0.05$, followed by a Gaussian
kernel with $\sigma=0.3T$. The occupancy was controlled using the
Broyden method \cite{broyden}. The self-energy was computed through
the ratio of correlators, $\langle \langle n_{\bar{\sigma}} d_\sigma;
d^\dag_\sigma \rangle\rangle / \langle \langle d_\sigma; d^\dag_\sigma
\rangle\rangle$ \cite{bulla}, corrected by the term $-w_\mathrm{UHB}/
\langle \langle d_\sigma; d^\dag_\sigma \rangle\rangle$, where
$w_\mathrm{UHB}$ is the spectral weight of the upper Hubbard peak
which was outside the NRG energy window (we redid some calculations
using the standard approach that explicitly includes the UHB in the
energy window, using a very large but finite value of $U$; we found
excellent agreement between the two computational schemes).

\section{Results  \label{sec-1}}

In this work we consider the temperature region $T \leq 0.02 D$,
which covers the range up to \unit[200]{K} if we assume $D\sim
\unit[10000]{K}$, i.e. $\O(1)$ eV. Here $D$ is the half bandwidth. We
study three densities (number of electrons per site) $n = 0.75, 0.8,
0.85$. These are typical of the over-doped and optimally doped
cuprates.

\subsection{DC Resistivity \label{resistivity}}

We begin with a summary of the results for the resistivity which form
the bedrock for this study. The findings in
\refdisp{badmetal,HFL,ECFL-resistivity} are extended in
\refdisp{high-T-resistivity} to higher temperatures, and from these we
have a fairly complete understanding of the behavior of $\rho$ at {\em
essentially all} $T$. A cartoon of these is sketched in
\figdisp{fig:schematic-rho}. The resistivity exhibits a variety of
dependences on $T$ upon warming from the absolute zero: (i) the
Gutzwiller correlated Fermi liquid (GCFL) regime with a quadratic T
dependence $\rho \propto T^2$ up to a (hole) density-dependent
Fermi-liquid temperature $T_{FL}(\delta)$ ($\delta=1-n$); (ii) the
Gutzwiller-correlated-strange metal (GCSM) regime with a linear T
dependence $\rho \propto T+ \mbox{constant} $ ($\mbox{constant} <0$),
(iii) a ``knee" connecting to the bad-metal (BM) regime with again a
linear $T$ dependence $\rho \propto T+ \mbox{constant} $
($\mbox{constant} >0$). This regime is so named since the $\rho$
crosses the fiduciary Ioffe-Regel-Mott maximal resistance $\rho_0$ at
temperature on the order of the Brinkman-Rice energy $\delta D$,
followed by (iv) a crossover to a high-temperature regime again with
linear $T$ dependence $\rho = A T$, devoid of an offset so that the line
extrapolates back to pass through the origin.

\begin{figure}[h]
  \centering
  \includegraphics[width=8cm]{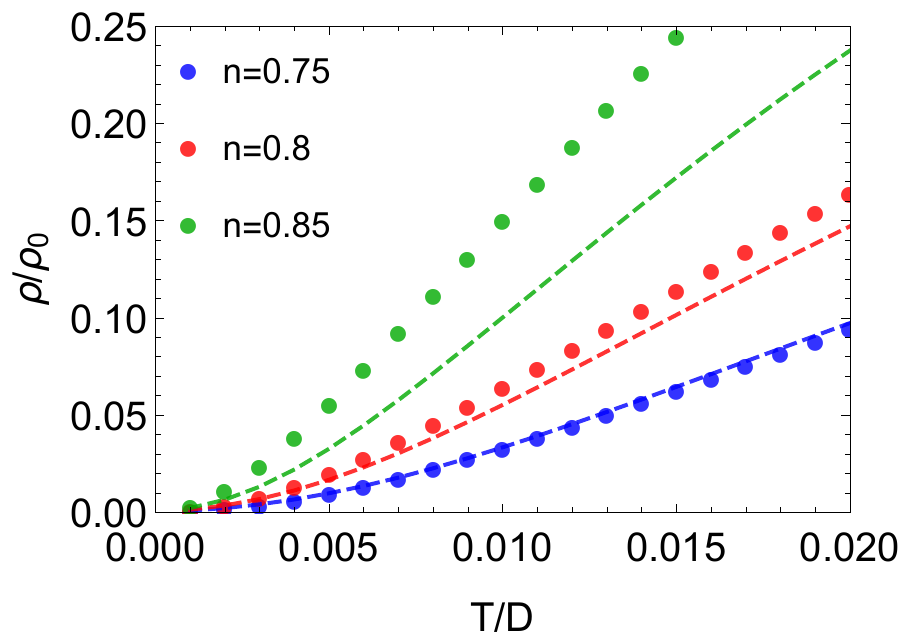}
    \caption{Comparison of the resistivity computed using the ECFL (symbols) and
  the DMFT (dashed). $\sigma_0=1/\rho_{0}$ is the Ioffe-Regel-Mott conductivity.
    As $n$ gets closer to unity,  the  ECFL scheme employed   systematically  underestimates $Z$ relative to the exact DMFT values  (see Fig.~6 of\refdisp{ECFL-resistivity}). This   lowers the effective Fermi temperature $T_{FL}$ and simultaneously enhances the magnitude of $\rho$  for $T>T_{FL}$, a feature that is  prominently visible above.  It should be possible to improve the quantitative agreement between the two theories in the future\cite{cutoff-comment}. }
  \label{fig:rho-compare}
\end{figure}

In \figdisp{fig:rho-compare} we
present the resistivity in the GCFL and GCSM regimes.
It is
striking that the GC strange metal has a robust linear $T$ resistivity over a
wide $T$ scale. The linear resistivity begins at $T_{FL}(\delta)$ which
can be driven to  low values, $\sim 45$K (see
\refdisp{ECFL-resistivity}), by the Gutzwiller correlations alone, even
though the bandwidth is of ${\cal O}(2)$ eV.  We emphasize that this unexpectedly drastic scale reduction yielding $T_{FL} \ll Z D \ll \delta D$ requires a ``hard'' calculation for justification and can hardly be argued from general principles.
The slight difference in the $T_{FL}(\delta)$ between the two theories
is due to the somewhat different $Z(\delta)$ found in the two theories, for example
Fig.~6 in \refdisp{ECFL-resistivity} shows that the ECFL gives a
smaller $Z$ than the DMFT\cite{cutoff-comment}.
We also note that using the standard value for  $\rho_{0} \sim 300 \; \mu \Omega$ cm, the
Ioffe-Regel-Mott resistivity \refdisp{Gurevitch-1981}, the absolute scale of the resistivity
computed in these approaches is quite similar to that found in the
experiments. For example, Fig.~1  in \refdisp{ECFL-resistivity}
compares well on an absolute scale with the well-known linear
resistivity result of S. Martin {\em et. al.} in \refdisp{Sam-Martin}
on Bi2212, where the superconducting phase cuts off the region $T \leq
80 $K.
\begin{figure*}[t]
  \centering
  \subfigure[]{\label{subfig:res_compare}\includegraphics[width=.65 \columnwidth]{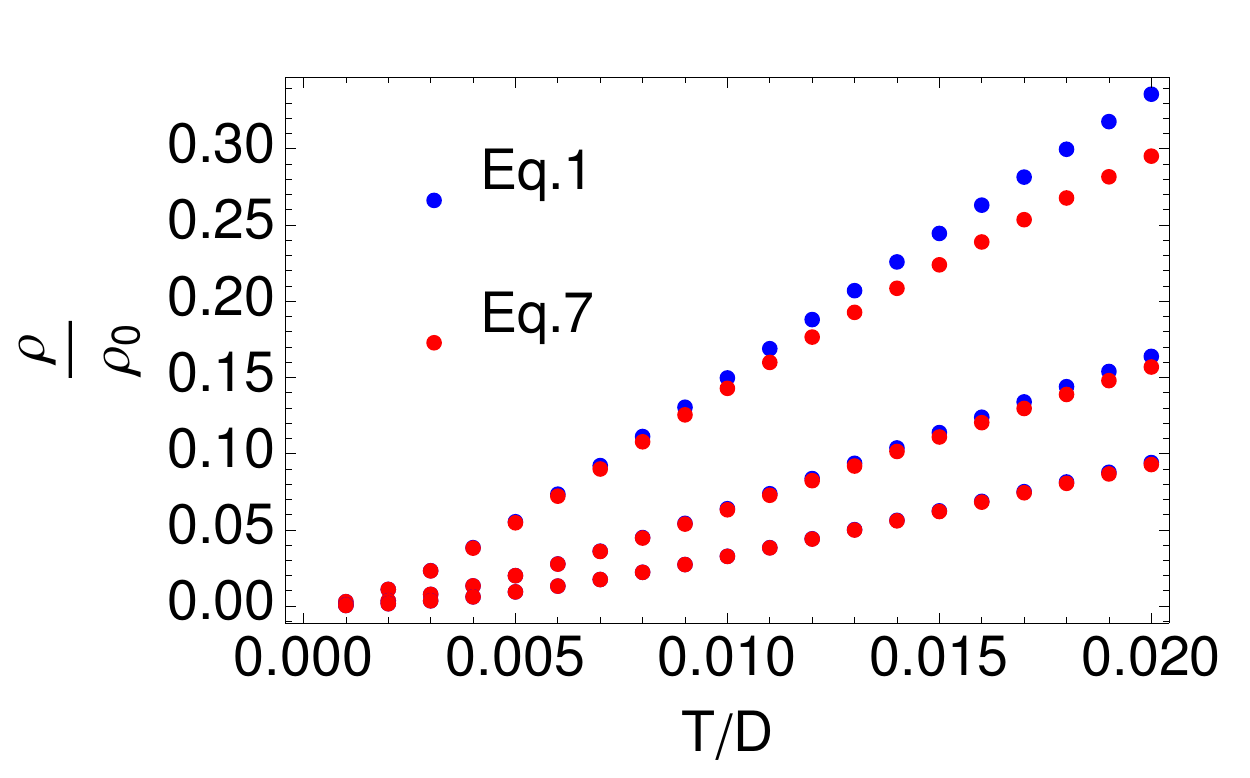}}
   \subfigure[]{\label{subfig:B0_B2}\includegraphics[width=.65 \columnwidth]{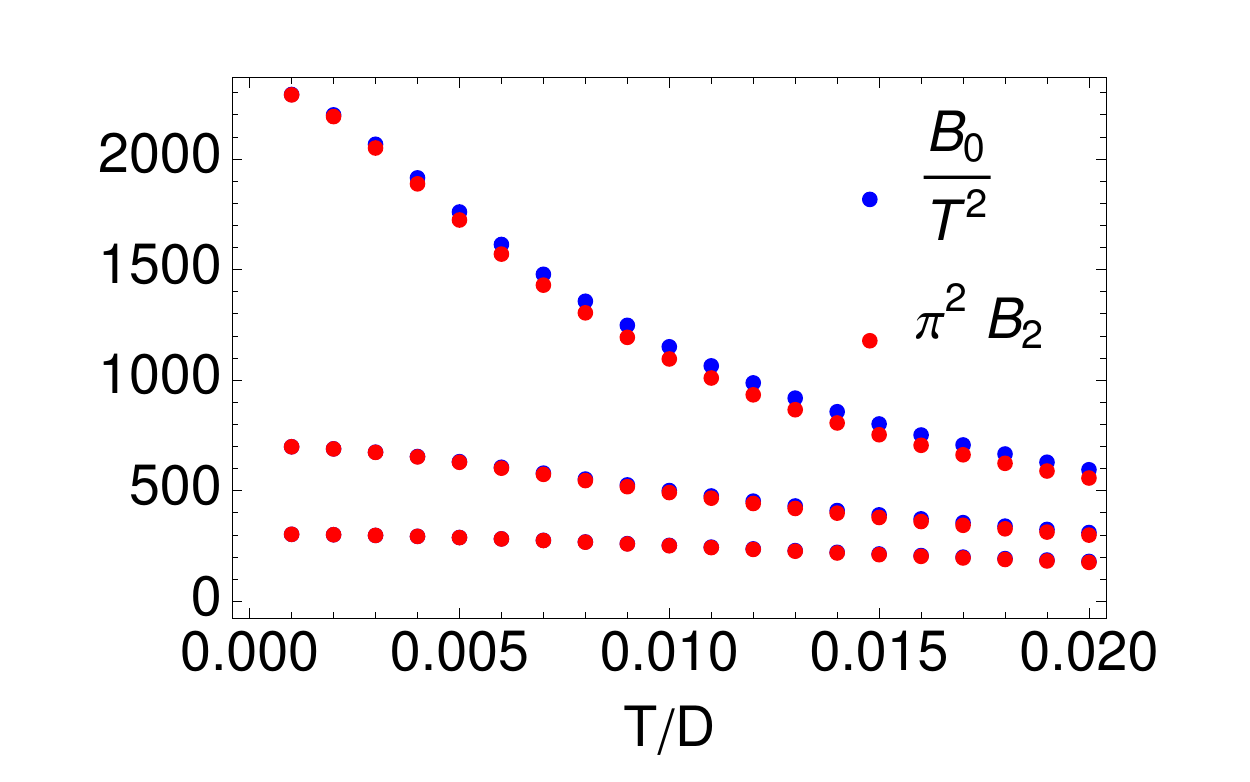}}
   \subfigure[]{\label{subfig:phi_A_0}\includegraphics[width=.65 \columnwidth]{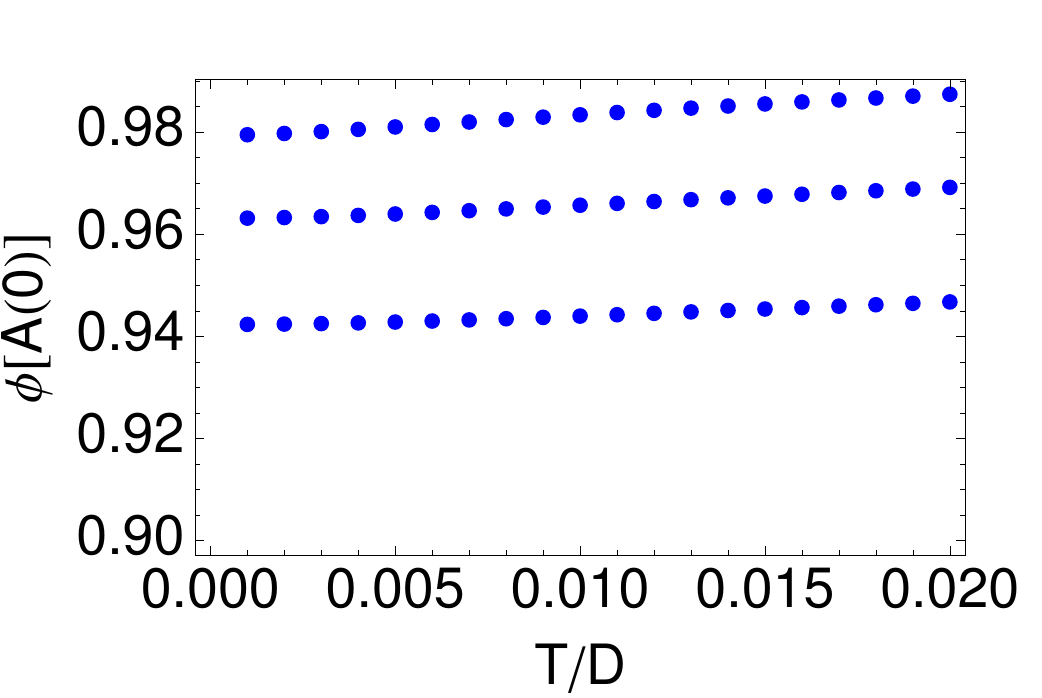}}
  \caption{ECFL calculation of the resistivity and related objects. {\bf Panel
  (a)}: The resistivity as a function of the temperature using the
  exact formula, \disp{conductivitydefn}, compared with the approximation, \disp{ressimple}, for $n=0.75,\ 0.8,\ 0.85$ (bottom to top). \disp{ressimple} is an excellent approximation at all densities for all temperatures.  {\bf Panel (b)}: Parameters resulting from a low-frequency expansion of the imaginary part of the self-energy in the vicinity of the Fermi-surace, plotted as a function of temperature, for $n=0.75,\ 0.8,\ 0.85$ (bottom to top). $B_0$ is the self-energy on the Fermi surface, while $B_2$ is the quadratic-frequency term. The ratio $\frac{B_2 \pi^2 T^2}{ B_0} \to 1$ as $T\to0$ and is approximately constant as a function of temperature. {\bf Panel (c)}: $\phi[A(0)]=\phi[\mu-\Re e \Sigma(0)]$, plotted as a function of the temperature, for $n=0.75,\ 0.8,\ 0.85$ (bottom to top). $\phi[A(0)]$ is practically independent of temperature, and has very weak density-dependence.
  }
  \label{fig:expansion_cond}
\end{figure*}

Building on the analysis of Refs. (\onlinecite{badmetal, HFL, ECFL-resistivity}), we derive a closed form expression for the resistivity in terms of the chemical potential and the real and imaginary parts of the single-particle self-energy on the Fermi surface [\disp{ressimple}]. We begin with the formula (Eq.~(41) in \refdisp{ECFL-resistivity}) for the conductivity on the infinite-dimensional Bethe lattice:
\beq
\sigma = 2 \pi D \ \sigma_0 \int d\omega \int d\epsilon \left(-\frac{\partial f}{\partial \omega}\right) \phi(\epsilon) \rho^2_G(\epsilon,\omega),
\label{conductivitydefn}
\eeq
where $\sigma_0 = e^2 \hbar \Phi(0)/D$    ($\Phi $ is defined in Eq.~(39) of \refdisp{ECFL-resistivity}), $ \sigma_0 = 1/\rho_{0}$, and the transport function
$\phi(\epsilon) =\Phi(\epsilon)/\Phi(0)$ is given explicitly in Eq.~(40) of \refdisp{ECFL-resistivity}
as $ \phi(\epsilon)= \Theta(1-\frac{\epsilon^2}{D^2}) \times (1-\frac{\epsilon^2}{D^2})^{\frac{3}{2}}$ .  The single-particle spectral function is
\beq
\rho_G(\epsilon,\omega) = \frac{1}{\pi} \frac{B(\omega)}{\left[A(\omega)-\epsilon\right]^2 + B^2(\omega)},
\eeq
where $A(\omega) \equiv \omega + \mu - \Re e \, \Sigma(\omega)$, $B(\omega) = - \Im  m \,\Sigma(\omega)$, and all objects depend implicitly on the temperature $T$. At low temperatures and frequencies  $B(\omega)\ll D $, so that  \disp{conductivitydefn} simplifies to
\beq
\sigma = \sigma_0 \int d\omega \left(-\frac{\partial f}{\partial \omega}\right) \frac{ \phi\left[A(\omega)\right]}{B(\omega)},
\label{sigsmallB}
\eeq
Following \cite{HFL}, we perform a small-frequency expansion
\beq
\phi\left[A(\omega)\right]= \phi\left[A(0)\right] + \ldots;\;\;\;\;\;\; B(\omega) = B_0 + B_2 \ \omega^2 + \ldots.
\label{frequencyexp}
\eeq
The linear order term in $B(\omega)$ as well as all higher order terms in $B(\omega)$ and $\phi\left[A(\omega)\right]$ make negligible contributions to the conductivity in the temperature range considered, and are therefore neglected. The integral may be evaluated analytically and yields
\beq
\si = \frac{ \si_0 \ \phi\left[A(0)\right] }{2 \pi T \sqrt{B_2 B_0}} \  \psi_1\left(\frac{1}{2}+\frac{1}{2\pi T}\sqrt{\frac{B_0}{B_2}}\right),
\label{condGamma}
\eeq
where $\psi_1(z)$ is the polygamma  function, related to the digamma function, $\Psi(z)$, through $\psi_1(z) \equiv \frac{d}{dz} \Psi(z)$ \cite{HilbertFermi}. The ratio $\frac{B_0}{B_2\pi^2T^2}$ is weakly dependent on temperature and may be replaced by its zero-temperature limit, see \figdisp{subfig:B0_B2}. In order to find this limiting value, consider  the GCFL regime where
\beq
B_0 =  B_2  \pi^2 T^2  \;\;\;\;(GCFL).
\label{Gamma0Gamma2GCFL}
\eeq
Substituting \disp{Gamma0Gamma2GCFL} into \disp{condGamma} and eliminating $B_2$, we finally obtain the simple formula
\beq
\rho = \frac{12  \ \rho_0 }{\pi^2  \ \phi\left[\bar{\mu}-\Re e \ \bar{\Sigma}(0)\right] }\times B_0,
\label{ressimple}
\eeq
where we have used that $\psi_1(1)=\frac{\pi^2}{6}$. Here, we  denote the zero-temperature limit of any variable $Q$ as $\bar{Q}$, and have used that $\phi\left[A(0)\right]$ is practically temperature independent [\figdisp{subfig:phi_A_0}]. Hence, the resistivity is proportional to the imaginary part of the self-energy on the Fermi surface. Moreover, the proportionality constant is very weakly density dependent (since this is true of $\phi\left[\bar{A}(0)\right]$). \disp{ressimple} can be obtained from Eq. (47) in \refdisp{ECFL-resistivity} by multiplying the RHS of the latter by the constant $\frac{12}{\pi^2}$ and setting $T\to0$ in the denominator. The latter equation is obtained by retaining the leading order term in the Sommerfeld expansion of \disp{sigsmallB}. In \figdisp{subfig:res_compare}, we plot the resistivity as a function of the temperature, using both Eqs. (\ref{conductivitydefn}) and (\ref{ressimple}), in the ECFL scheme. We find that \disp{ressimple} is an excellent approximation at all densities and temperatures considered, i.e., it holds in both the GCFL and GCSM regimes. 

In the GCFL regime, substituting \disp{Gamma0Gamma2GCFL} into \disp{ressimple}, and using the fact that $B_2$ is approximately constant, we find that
\beq
\rho = \frac{ 12  \bar{B}_2 \ \rho_0}{\phi\left[\bar{A}(0)\right] }\times T^2 \;\;\;\;\;(GCFL).
\label{resGCFL}
\eeq
From Fig. 7 of \refdisp{ECFL-resistivity}, we know that $\bar{B}_2\propto \frac{1}{\bar{Z}^2}$, where $Z$ is the quasiparticle weight on the Fermi surface. Therefore, \disp{resGCFL} implies that $\rho\propto \frac{T^2}{\bar{Z}^2}$ in the GCFL regime.

\begin{figure}[h]
\centering
\includegraphics[width= .95\columnwidth]{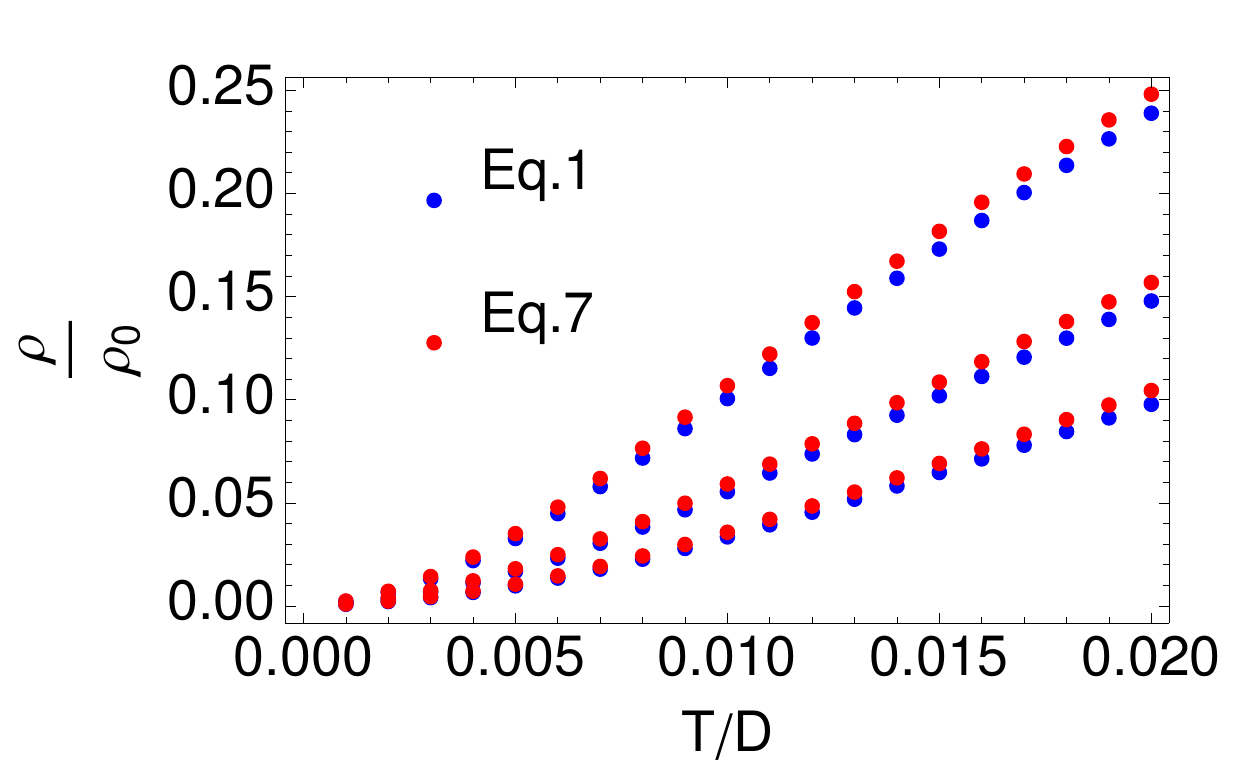}
\caption{The exact resistivity [\disp{conductivitydefn}] compared with the approximation \disp{ressimple}, using the DMFT calculation for $n=0.75,\ 0.8,\ 0.85$ (bottom to top). \disp{ressimple} is an excellent approximation at all densities for all temperatures. [See \figdisp{subfig:res_compare} for the corresponding figure in ECFL.]}
\label{DMFT_Res}
\end{figure}

In Fig. \ref{DMFT_Res}, we plot the exact resistivity, together with
the approximation \disp{ressimple}, both
obtained using the DMFT calculation [corresponding to \figdisp{subfig:res_compare} in the case of ECFL]. Once again, we find that \disp{ressimple}
is an excellent approximation at all densities and temperatures considered, i.e., it holds in both the GCFL and GCSM regimes.

Finally, we note that the important effective Fermi temperature, $T_{FL}$, can be estimated as the temperature at which  the resistivity deviates from its low-temperature quadratic behavior. We find at the three densities considered, the so-determined effective Fermi temperature for ECFL is,  in agreement with \refdisp{ECFL-resistivity}, given by $T_{FL}\sim .05 \bar{Z} D$. In the case of DMFT, we also find $T_{FL}\sim .05 \bar{Z} D$, where a slightly higher value of $\bar{Z}$ results in a slightly higher value of $T_{FL}$, as compared to ECFL.

%

\subsection{Chemical potential and compressibility \label{compressibility}}

The chemical potential in the ECFL theory is found from the
self-consistency condition of the Green's function. The
compressibility $\kappa = n^{-2} \pt n / \pt \mu$ is determined by
numerical differentiation. 
The derivative is computed using the finite difference formula $\pt
n/\pt \mu = [(n+\delta n)-n]/[\mu(n+\delta n)-\mu(n)]$ with $\delta
n=0.001$. In the DMFT we used larger $\delta n=0.01$ and we performed
two full DMFT runs for fillings $n$ and $n+\delta n$.

\begin{figure}[h]
  \centering
  \includegraphics[width=.64 \columnwidth]{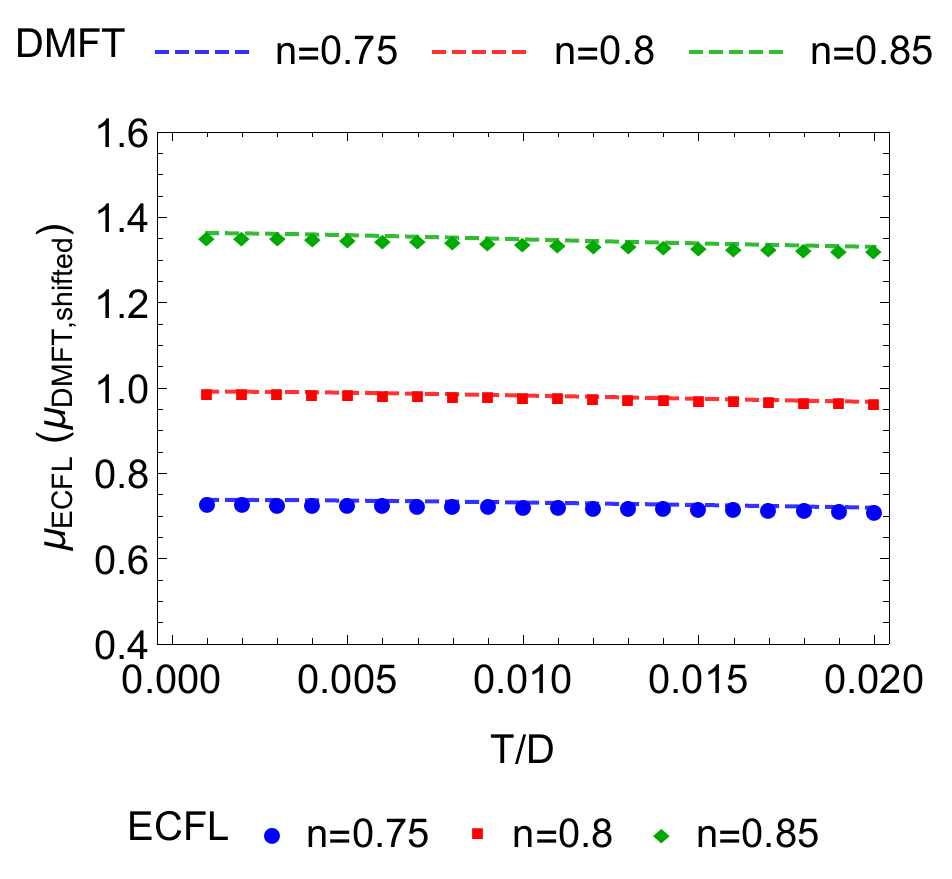}
  \caption{Chemical potentials at $n = 0.75,\ 0.8,\ 0.85$ for ECFL (symbols) and DMFT (dashed lines). The DMFT results 
    are shifted by a density-dependent constant. After the
  shift, the chemical potentials almost coincide.}
  \label{fig:mu}
\end{figure}

We see that the chemical potentials (Fig.~\ref{fig:mu}) match well
apart from a constant shift \cite{chemical-shift}. The results
obtained using two different impurity solvers (NRG and CT-HYB QMC) in
the DMFT are in agreement, thus the difference is not related to some
technical issue in the NRG, but is an actual discrepancy between DMFT
and ECFL.

In our earlier work on the single impurity Anderson model
\cite{Edward-Sriram-Alex}, using a scheme that is an adaptation of
that in \refdisp{ECFL-DMFT}, we studied the single impurity energy,
which is a close analog of the chemical potential in the present
problem. There we found that the location of the impurity energy found
from the second order ECFL equations matched very closely the impurity
energy found in the NRG (see Table 1 in \refdisp{Edward-Sriram-Alex}). In
view of that excellent agreement, the current discrepancy on the
absolute scale of the chemical potential between the DMFT results
(also from NRG) and the present second order scheme is somewhat
unexpected. It would appear that the different hole number sum rule
and the cutoff scheme used here relative to the scheme in
\refdisp{Edward-Sriram-Alex,ECFL-DMFT} influences this variable- and
needs to be investigated more closely in the future.

\begin{figure}[h]
  \centering
  \subfigure[]{\label{subfig:compressibility}\includegraphics[width=.8 \columnwidth]{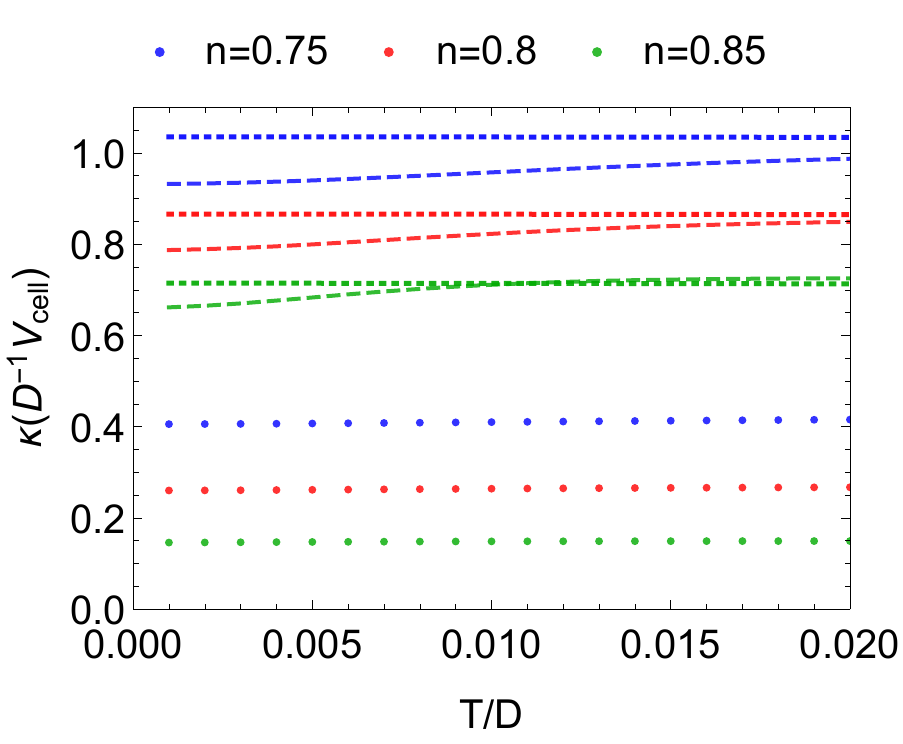}}
   \subfigure[]{\label{subfig:Z_kappa}\includegraphics[width=.8 \columnwidth]{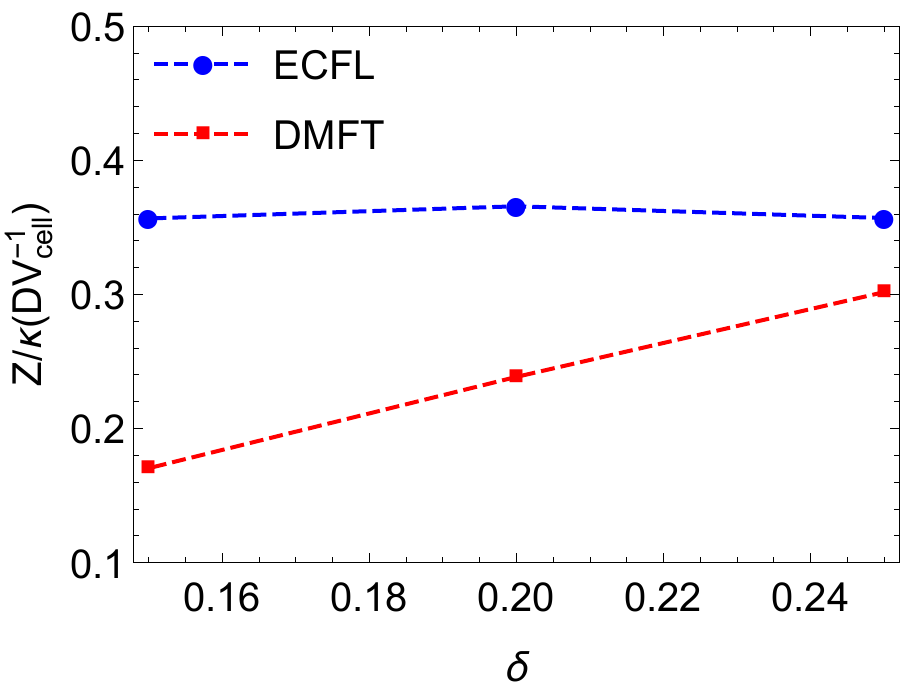}}
   \caption{(a) Compressibility $\kappa = n^{-2} \pt n / \pt \mu$ of ECFL (symbols), DMFT (dashed lines) and free fermions (dotted lines). 
     The DMFT results give a systematically higher value of compressibility than the ECFL theory.
(b) $Z/\kappa$ for the lowest temperature $T=0.001D$ at the three densities considered for ECFL (blue) and DMFT (red). The ECFL result for the compressibility is proportional to the quasiparticle weight Z, unlike the DMFT result which displays some variation. The difference in compressibility  between  the two theories seems related to the density dependent shift in chemical potentials noted in \figdisp{fig:mu}. }
  \label{fig:compressibility}
\end{figure}

We note that the compressibilities (Fig.~\ref{fig:compressibility})
are also roughly similar, and both theories show a suppression
relative to the free fermion theory. The free fermion theory shows a
slight monotonic decrease of the compressibility with $T$. In the GCFL
and GCSM regimes, the ECFL compressibility shows an increase with $T$,
followed by a slight fall with $T$ in the bad metal regime. In
Fig. \ref{subfig:Z_kappa}, we show that in the ECFL theory $Z/\kappa$ is
a constant within numerical errors ($\sim \pm 3.4\%$) at $T = 0.001D$.
This is not the case in the DMFT, where $Z$ is proportional to
$\delta$, while $\kappa$ behaves approximately as $\kappa \propto
\delta^{0.2}$ close to the doping-driven Mott transition
\cite{ECFL-DMFT}.
In the GCFL regime, if we assume that the limit $n\to 1$ follows the
almost localized Fermi liquid
theory\cite{Vollhardt-RMP,Vollhardt-Anderson}, we should expect the
compressibility to scale with $Z$. This is in accord with the results
of ECFL \figdisp{fig:compressibility} panel (b) but not with the
DMFT.


\subsection{Bubble Susceptibility \label{bubbles}}
The knowledge of the Green's functions and the numerically determined exact compressibility and  magnetic susceptibility $\chi_{spin}$ [see below Sec. \ref{susceptibility}] enable us to check a popular assumption of retaining only the bubble graphs, and throwing away the vertex correction for these quantities.  We write the charge susceptibility $\chi_c=dn/d\mu$ as
\barray
\chi_{c}&=&   \frac{1}{\beta N_s} \, \frac{d}{d \mu
}\,\sum_{k,\omega_n,\sigma} e^{i \omega_n 0^+} G_\sigma(k,i\omega_n)\nn \\
&=& -   \frac{1}{\beta N_s}\sum_{k,\omega_n,\sigma}  G^2_\sigma(k,i\omega_n) \{1 -  \frac{d}{d \mu } \Sigma_\sigma(k,i \omega_n)\}\;
\earray
and similarly for $\chi_{spin}$ by replacing $\frac{d}{d \mu }\to \frac{d}{d B }$, where $B$ is the magnetic field. The vertex corrections thus  correspond to the $\mu$ or $B$ derivatives  of the {\em self energy}.
Approximating this by dropping the derivative of the self energy, we get $\chi_c \sim \chi_{spin} \sim \chi_{Bubble}$ where
\barray
\chi_{Bubble}
&=& -   \frac{1}{\beta N_s }\sum_{k,\omega_n,\sigma}  G^2_\sigma(k,i\omega_n). \;
\earray
  As usual  we can  convert the sum to a contour integral using the pole structure of the Fermi function $f(\omega)$ and  write 
\barray
\chi_{Bubble}&=& \frac{2}{N_s} \sum_k \int_{\Gamma} \frac{d\omega}{2 \pi i} f(\omega) G^2(k,\omega) \nn \\
&=& \frac{2}{\pi N_s} \sum_k \int d \omega f(\omega) \Im  m \, G^2(k, \omega+ i 0^+), \label{bubble1}
\earray
where $\Gamma$ is a closed contour encircling the imaginary axis in a
counterclockwise fashion, and we rotated the axis to a pair of lines parallel to the real axis to obtain the final line. Using the standard definition of the spectral function $\rho_G(k, \omega)= - \frac{1}{\pi} \Im m \, G(k, \omega+ i 0^+)$
we may write $ \Im  m \, G^2(k, \omega+ i 0^+) = (- 2\pi)  \Re e\,
G(k, \omega) \; \rho_G(k, \omega)$ to express $\chi_{Bubble}=  -
\frac{4}{N_s} \sum_k \int d \omega f(\omega) \Re e \, G(k,
\omega)\rho_G(k, \omega)$.  In the limit $d\to \infty$  the Dyson self energy is independent of $k$, and therefore we can write $ \Im  m \, G^2( \epsilon , \omega+ i 0^+) =  \Im  m \, \frac{d}{d \epsilon} G(  \epsilon, \omega+ i 0^+) = - \pi  \frac{d}{d \epsilon} \rho_G(\epsilon,\omega)$, where we exchanged the two operations in the last line. Using the definition of the single particle momentum distribution function $n_k \to n(\epsilon) \equiv \int d\omega f(\omega) \rho_G(\epsilon,\omega)$ we can perform the $\omega$ integration in \disp{bubble1} and get a compact relation valid in high dimensions:
\beq
\chi_{Bubble} =- 2 \int d\epsilon\, {\cal D}(\epsilon) \; \frac{d}{d \epsilon} n(\epsilon) \label{bubble2}.
\eeq 
Here ${\cal D}(\epsilon) = \frac{2}{\pi D} \sqrt{1- \epsilon^2/D^2}$
is the band density of states per site per spin, and $D$ is the half
bandwidth. 

For noninteracting electrons the function $n(\epsilon)$ is
a constant with a unit jump  at $\epsilon_F$, and we recover the standard result $\chi^0= 2 {\cal D}(\epsilon_F)$. 

\begin{figure}[h]
  \centering
  \subfigure[]{\label{subfig:nofelow}\includegraphics[width=.8 \columnwidth]{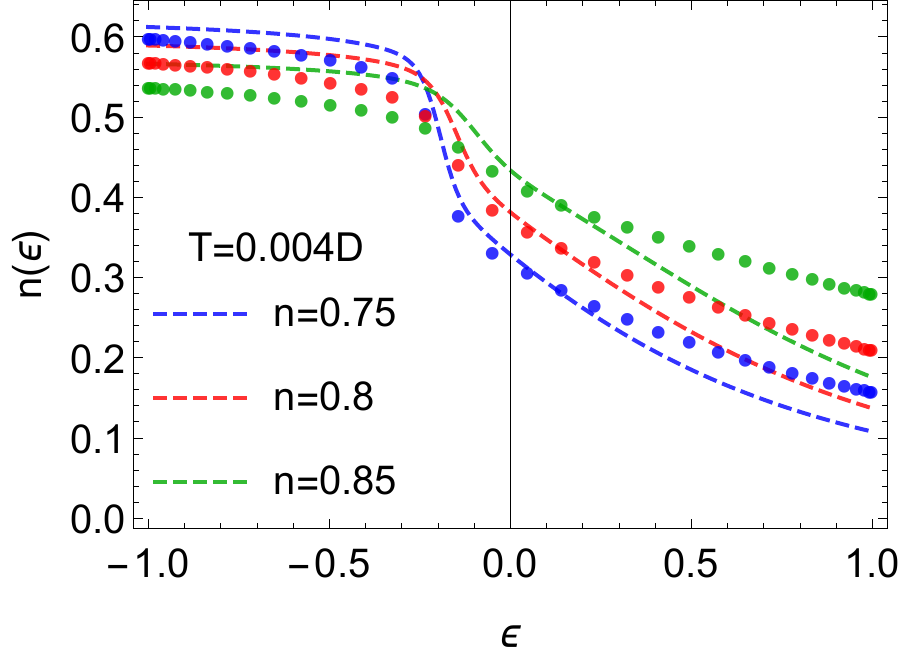}}
   \subfigure[]{\label{subfig:nofehigh}\includegraphics[width=.8 \columnwidth]{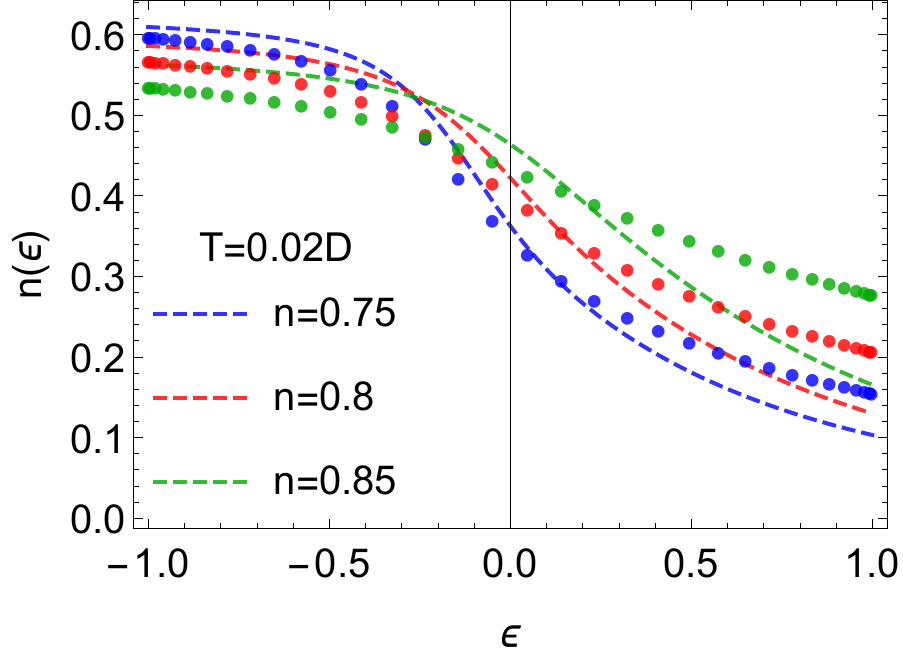}}
   \caption{The momentum distribution curves at three densities
   $n=.75,.8,.85$ (top to bottom at $\epsilon=-1$) at  T=.004 D [panel (a)] and T=.02 D [panel (b)]. The ECFL curves are solid symbols and the DMFT curves are dashed lines. 
   }
  \label{fig:nofe}
\end{figure}

\begin{figure}[h]
  \centering  
 \subfigure[]{\includegraphics[width= .8\columnwidth]{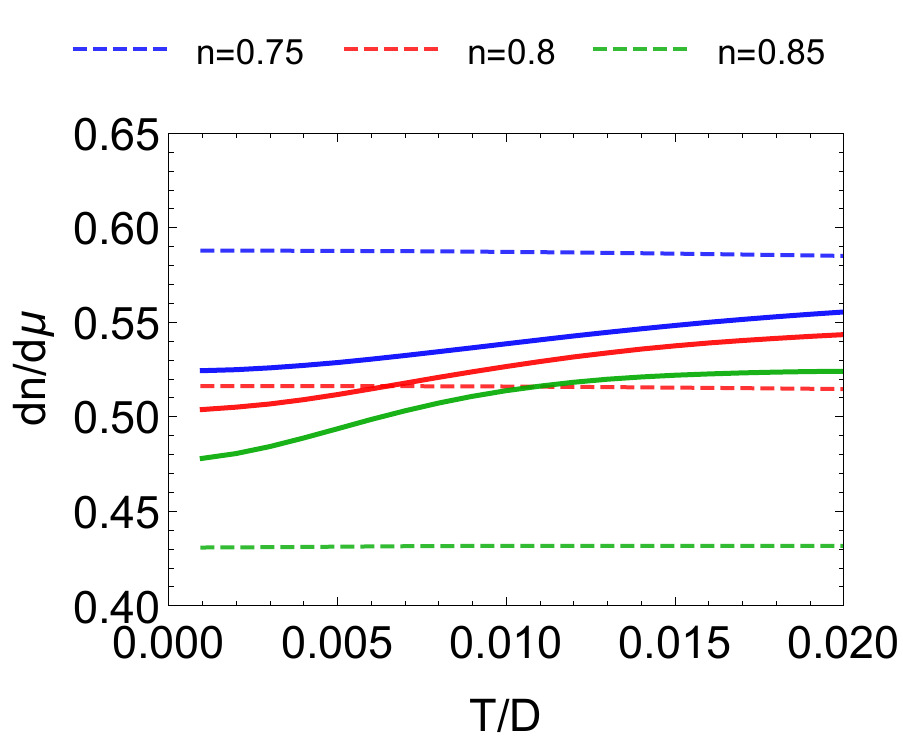}}
 \subfigure[]{\includegraphics[width= .8 \columnwidth]{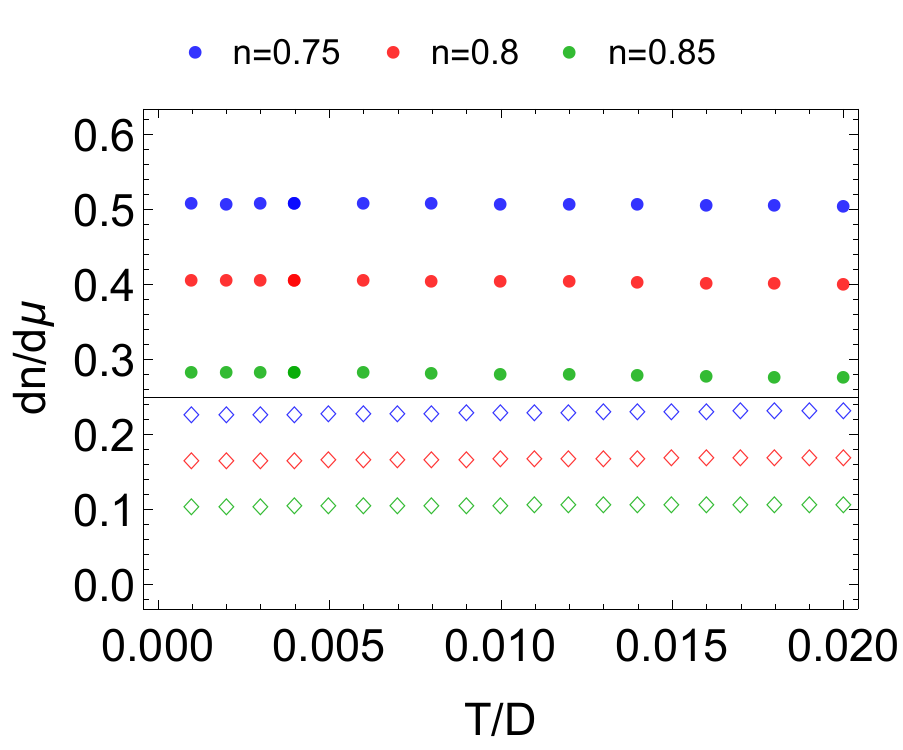}}
\caption{ The   charge susceptibilities $\chi_c = dn/d\mu$, which are related to compressibility $\kappa$ as $ \chi_c=n^2 \kappa$. The numerically exact values  versus  bubble estimates [\disp{bubble2}] in panel (a) DMFT (full and  dashed lines) and in panel (b) from ECFL (empty diamonds and solid circles). }
\label{fig:bubble-chi}
\end{figure}

In the correlated problem, the jump at the Fermi energy is $Z_k$ by
Migdal's theorem, and so its contribution to $\chi_{Bubble}$ is $Z_k$.
The background also contributes to the integral in \disp{bubble2}, 
and it is important to understand its behavior as $n \to 1$.  In
\figdisp{fig:nofe} we display the momentum distribution at the three
densities considered at two temperatures. We note that the entire
variation of the monotonic function $n(\epsilon)$ is on the scale of
$\delta$; it settles down to a flat function $n(\epsilon)=0.5$ {\em
at} $n=1^-$ and for small departures from half filling, the occupied
(unoccupied) region is enhanced (depleted) by an area that is
proportional to $\delta=1-n$. Thus we see that as $n \to 1$, the
background contribution is at most as large as $\delta$, and thus
$\chi_{Bubble}$ is a suitably weighted average of $\delta$ and $Z$. 
In the density regimes we are considering, the $\delta$ variation of
$Z$ is close to $\delta^{1.39}$ rather than $\delta$ (see discussion
in \refdisp{ECFL-DMFT}), and hence this balance can only be determined
by a numerical evaluation. From \disp{bubble2} we can evaluate
$\chi_{Bubble}$, and the results are shown from both theories at the
three densities $\delta=.25,.2,.15$ in \figdisp{fig:bubble-chi}.
Within ECFL it appears that $\chi_{Bubble} $ is dominated by the
Migdal jump contribution; the spacing between the three relatively
constant lines increases at lower $\delta$. Within DMFT the situation
appears to be reversed and $\chi_{Bubble}$ seems to scale with
$\delta$.  
In \figdisp{fig:compressibility} we see that the DMFT results for
$Z/\kappa$ have a distinct positive slope relative to the ECFL
results, and this is consistent with the above discussed differences
in the computed $\chi_{Bubble}$ as well.

\subsection{Self-energy and local density of states \label{selfenergy}}

\begin{figure}[h]
  \centering
  \subfigure[]{\label{subfig:sigma-n1}\includegraphics[width=.64 \columnwidth]{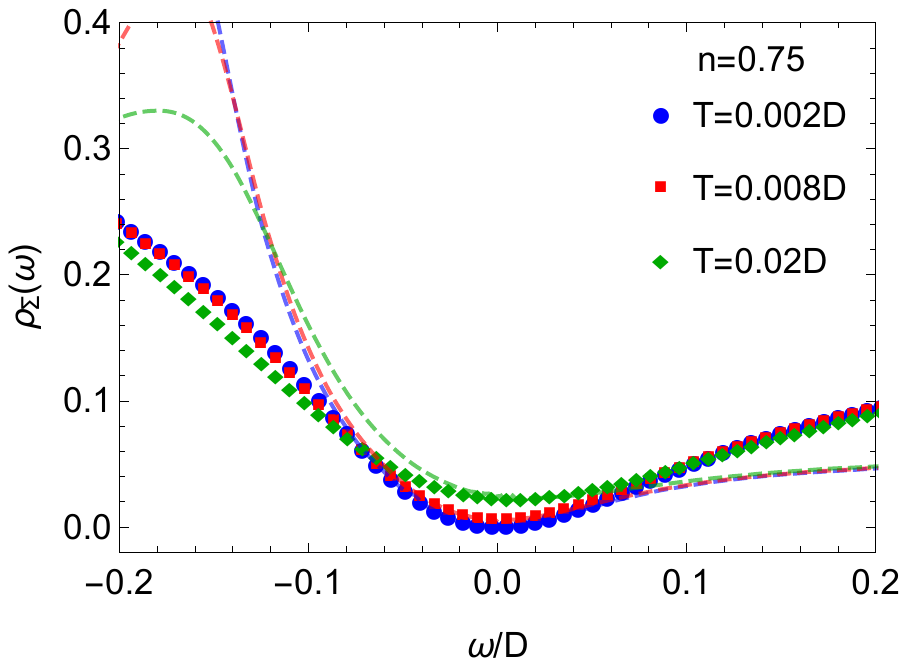}}
   \subfigure[]{\label{subfig:sigma-n2}\includegraphics[width=.64 \columnwidth]{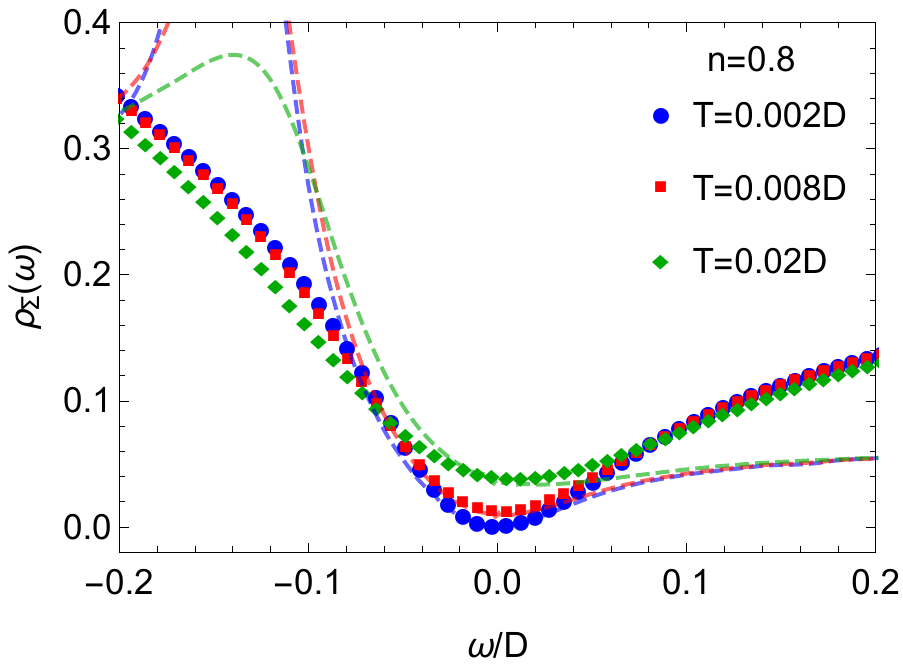}}
   \subfigure[]{\label{subfig:sigma-n3}\includegraphics[width=.64 \columnwidth]{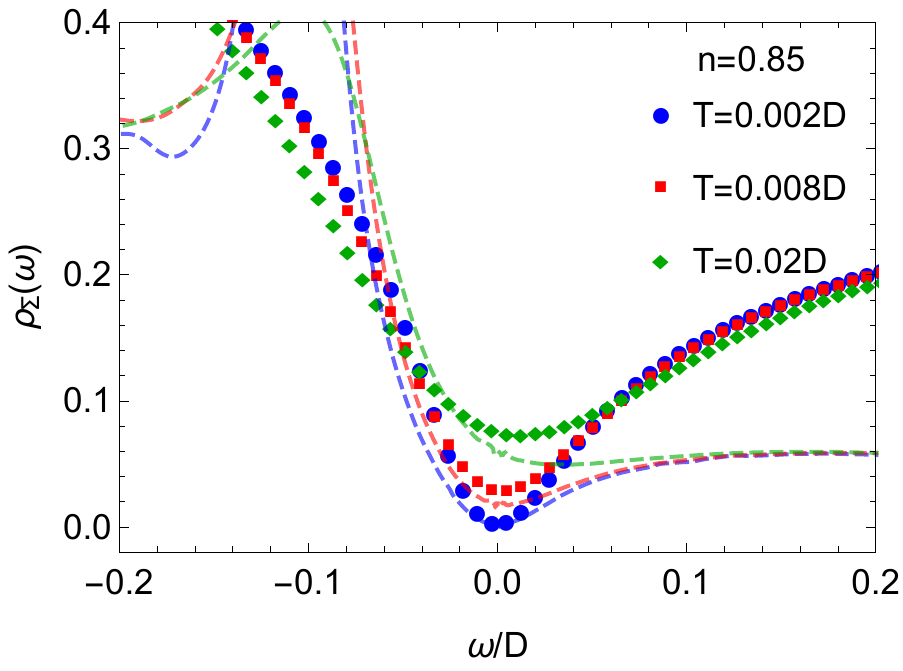}}
  \caption{Single particle decay rates, i.e. the spectral functions of self-energy [$\rho_\Sigma(\omega) =
  -\pi^{-1} \Im m \Sigma(\omega)$] of ECFL (symbols) and
  DMFT (dashed lines) for a range of temperatures.}
  \label{fig:sigma}
\end{figure}

In this section we study  the imaginary part of the  self energy $\rho_{\Sigma}(\omega)= -\frac{1}{\pi} \Im m \Sigma(\omega)$ and the (local) spectral
function integrated over the band energies  $\rho^{loc}_{G}(\omega)= -\frac{1}{\pi} \Im m \int d \epsilon \;{\cal D}(\epsilon)\, G(\epsilon,\omega)$.
The results of the two theories, including the magnitudes and their
variation,   are very close  at  low energies.  The ECFL self-energy misses a
maximum  in $\rho_{\Sigma}(\omega)$ found in DMFT between $\omega\sim -0.1D$ and $\omega \sim - 0.2D$, see Fig. \ref{fig:sigma}. This feature was already noted in \refdisp{ECFL-DMFT} and it is
expected to influence the results of various quantities, such as the
optical conductivity and dynamical Hall constant, but only at a fairly
large energy.   The imaginary part of the self energy in both theories shows a significant $\omega^3$ type (i.e., odd in frequency) correction to the simple-minded expectation of a $\omega^2$ behavior  from Fermi liquid theory. This type of a skew has been argued in \refdisp{ECFL-ARPES} to be responsible for the  unusual and distinctive spectral functions in real materials- such as the cuprates.

\begin{figure}[h]
  \centering
  \includegraphics[width=.75 \columnwidth]{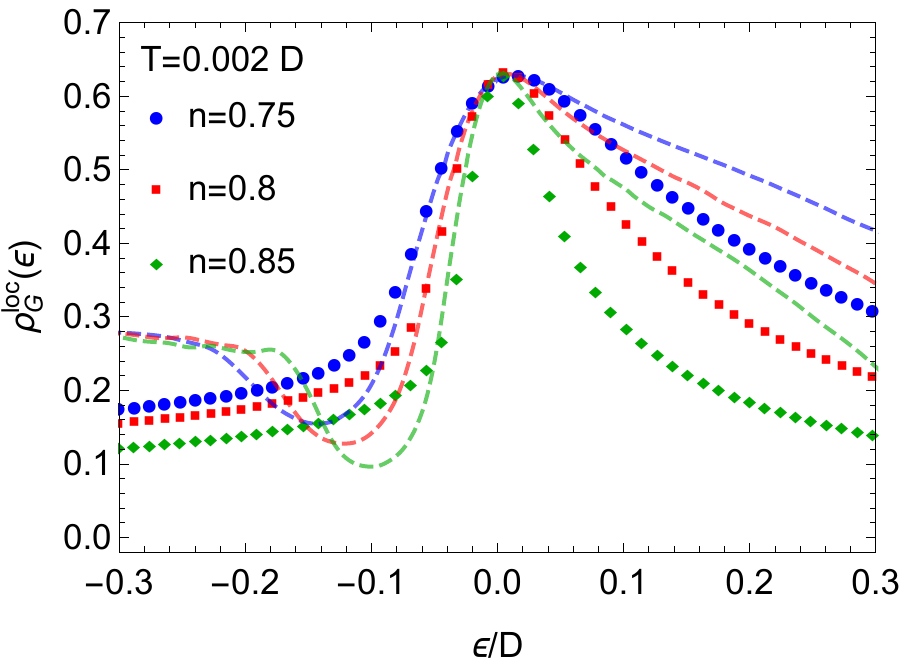}
  \caption{Local density of states $\rho^{local}_{G}(\epsilon)$ of ECFL (symbols) and DMFT (dashed lines) at $T=0.002D$.  }
  \label{fig:rho}
\end{figure}

The local spectral functions of the two theories, shown in Fig.~\ref{fig:rho}, are similar. They exhibit a sharpening of the maximum as $n$ increases.  Let us note that this object is relevant for angle integrated photoemission studies as well as STM studies, where one would also have to correct for the one electron density of states showing structure beyond that in the present theory.



\subsection{Entropy and  heat capacity \label{heatcapacity}}
The heat capacity is computed in the ECFL theory by numerically
differentiating the internal energy
as $C_V = \pt
E_K/\pt T$ on a fine $T$ grid. From its numerical integration  $\int^T_0 dT'
C_V(T')/T'$ we find the entropy. A similar procedure is used in the DMFT: The kinetic energies
were computed on an equally spaced temperature grid (step size $\Delta
T=10^{-3}D$), numerically differentiated, smoothed using a Gaussian
filter to obtain the heat capacity $C_V$, then interpolated using 
second-order polynomials, and finally integrated to obtain the entropy.

\begin{figure}[h]
  \centering
   \subfigure[]{\label{subfig:Cv}\includegraphics[width=.63 \columnwidth]{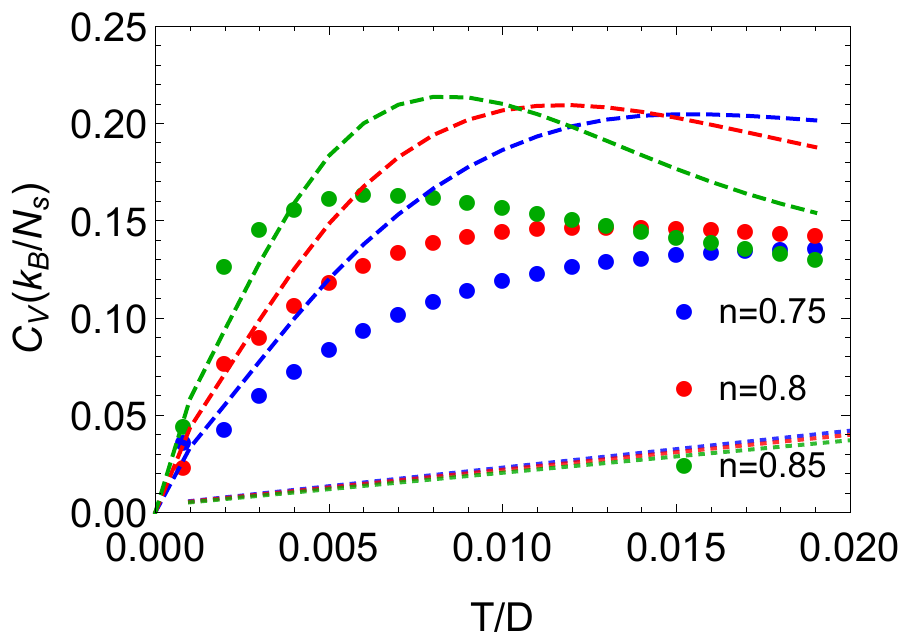}}
    \subfigure[]{\label{subfig:CvT}\includegraphics[width=.63 \columnwidth]{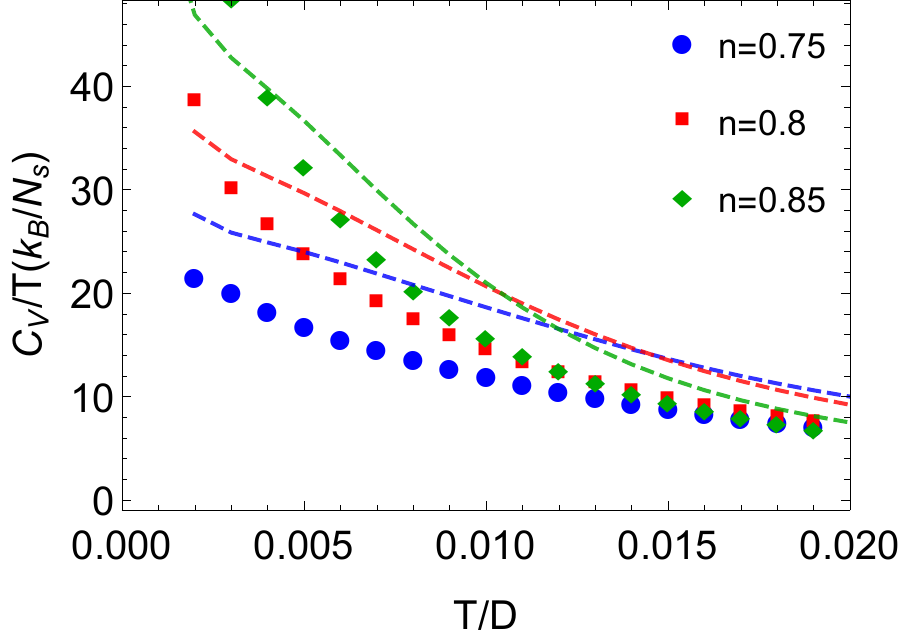}}
    \subfigure[]{\label{subfig:gammaZ}\includegraphics[width=.63 \columnwidth]{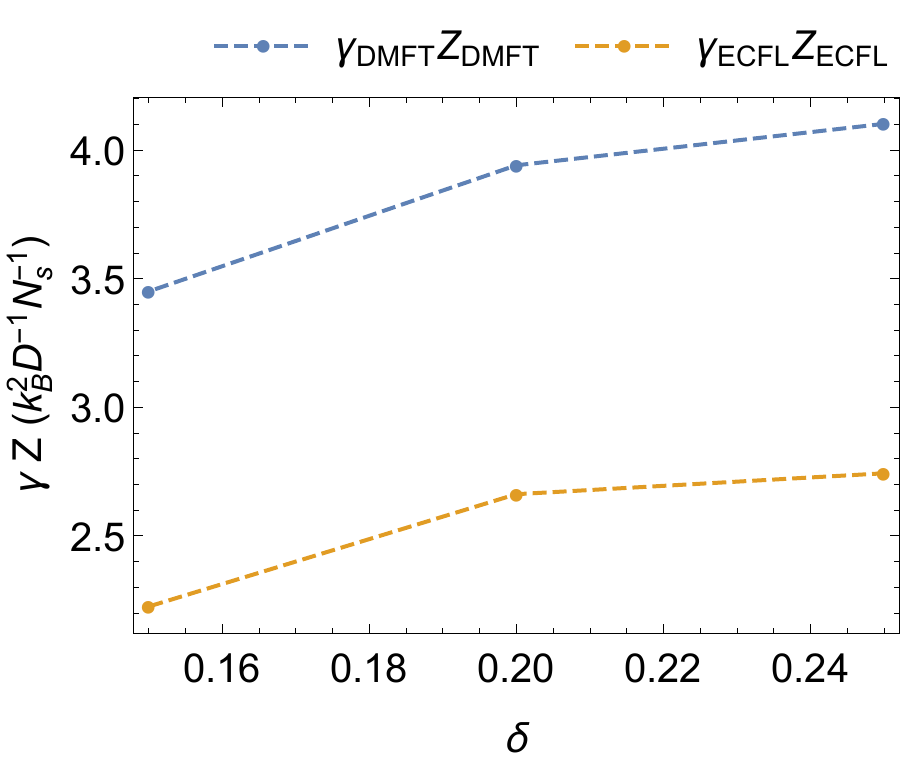}}
  \caption{(a) Specific heat computed from the kinetic energy by differentiation as $C_V = \pt E_K/\pt T$ for ECFL (symbols), DMFT (dashed lines) and free fermions (dotted lines). For $n=0.8$ and $n=0.85$  the heat capacity shows a gentle  maximum at  a characteristic $T$.  (b) The ratio
  $C_V/T$ versus $T$ of ECFL (symbols) and DMFT (dashed lines). Taking the ratio with $T$ wipes out the maximum seen in (a).
  (c)
  $\gamma \times Z$ at $T=0.001D$.
  }
  \label{fig:entropy-Cv}
\end{figure}

The heat capacity $C_V$ is displayed in Fig.~\ref{subfig:Cv}. We note that $C_V$  has a Schottky peak near $T\sim T_{FL}$ which
becomes sharper as the density increases. At lower densities ($n=0.7$,
$0.75$), a linear-$T$ behavior is resolved, as we expect for a Fermi
liquid. In Fig.~\ref{subfig:CvT} we display $C_V/T$, from which we see that  for  densities closer to half-filling
($n=0.8,0.85$), the linear behavior of heat capacity is not clearly resolved due to the small $T_{FL}$
scale, and also due to increasing numerical uncertainties near half
filling. Consequently, we find $C_V/T$ appears to be growing as $T$
decreases, instead of saturating.  In Fig.~\ref{subfig:gammaZ} we show the product of the
heat-capacity slope $\gamma=C_V(T)/T$ and the quasiparticle residue
$Z$ at a low $T$ corresponding to the GCFL regime.  This product is expected to be a constant for localized Fermi liquids
\refdisp{Vollhardt-RMP}. At $\delta = 0.15$, we see
however some variation in both ECFL and DMFT results. For higher hole densities $\delta = 0.2,
0.25$
, it is indeed almost a constant.
\begin{figure}
  \centering
 \subfigure[]{\label{subfig:entropy}\includegraphics[width=\columnwidth]{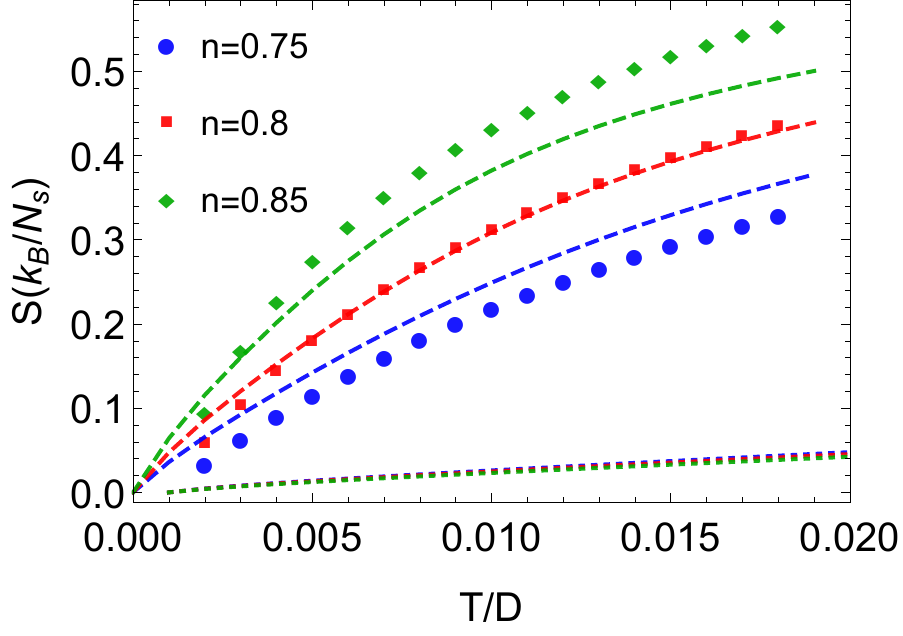}}
%
\caption{ The entropy versus $T$ computed as $\int^T_0 dT' C_V(T')/T'$ for ECFL (symbols), DMFT (dashed lines), and free fermions (dotted lines).
 }
%
\label{fig:entropy}
\end{figure}

  In Fig.~\ref{fig:entropy} we plot the entropy of the two theories, which give very similar results,  and that of the free Fermi gas with a much lower entropy recovery at these temperatures.
It is revealing to compare the heat capacity curve at $n=0.8$ in \figdisp{subfig:Cv}, with the
resistivity results in \figdisp{fig:rho-compare} at the same
densities. Both theories show a broad maximum in the heat
capacity near the corresponding Fermi liquid temperature
$T_{FL}(\delta)$; this is  the temperature where the  GCFL quadratic
behavior of resistivity gives way to a linear behavior of the GCSM. At this
temperature the entropy per site [see \figdisp{fig:entropy}] is $\sim 0.2 \, k_B$, compared to the high $T$ ($T=\infty$) value
of $1.0119\; k_B$, obtained from  $S_{ideal}\equiv S_{T=\infty}=
 k_B \left\{  n \log 2 - n \log n - (1-n) \log(1-n) \right\}$.   This
corresponds to about 20\% release of the entropy. For comparison, the Fermi gas on the Bethe lattice  releases much less, about {1-2}\% entropy at  a comparable $T/D$.  At lower particle densities $n=0.8,
0.75$ we again see that a $\sim 15-20 \%$ release of the entropy
occurs at the corresponding Fermi liquid temperature $T_{FL}(\delta)$,
however the heat capacity has a more rounded behavior.

\begin{figure}
  \centering
  \includegraphics[width= \columnwidth]{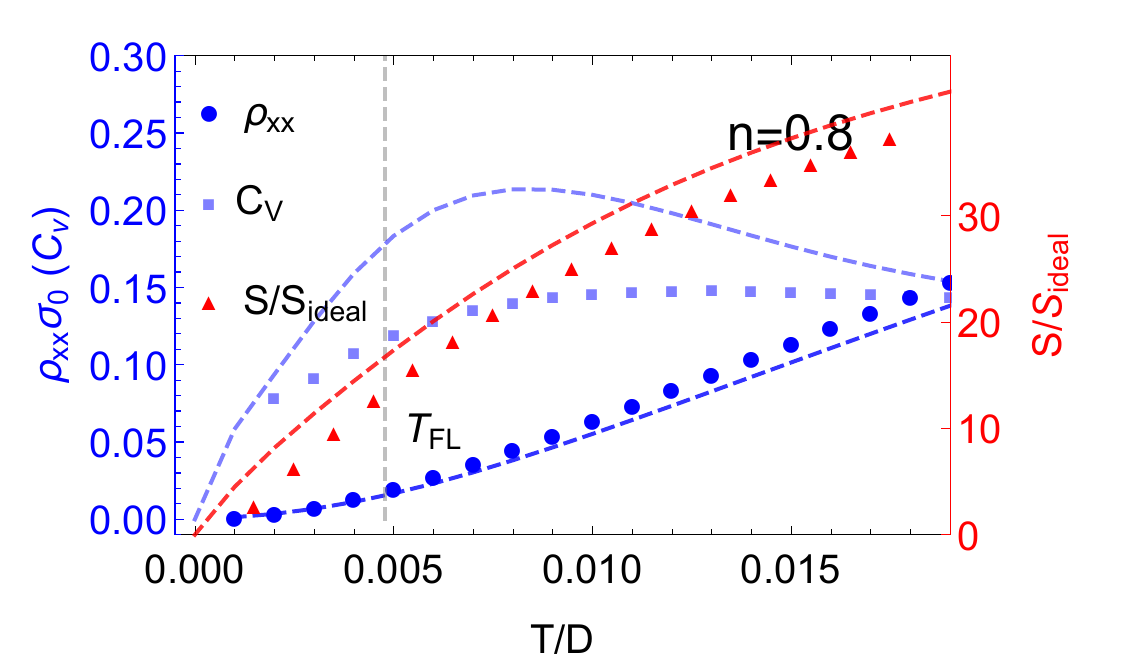}
\caption{ Resistivity (blue
  circle), specific heat (light blue square),
  and entropy ({red triangle}) as percentage of the
ideal entropy at infinite temperature $S_{ideal}$.  The (Schottky) peak in the heat capacity is close to $T_{FL}$, the onset point of the linear-$T$ resistivity,
or the end of the crossover region.}
\label{fig:rho-xx-Cv-S-kappa-n085}
\end{figure}

In order to explore this further, in
\figdisp{fig:rho-xx-Cv-S-kappa-n085} we display the resistivity and
the entropy recovery on the same $T$ scale. We may thus take as a rule
of thumb that at $T_{FL}$, the GCFL   entropy release is
$\sim 15-20 \%$ relative to the maximum. This implies a substantial   loss of coherence relative to the Fermi gas, i.e., the disordering of either the configurational  (i.e., charge) degrees of freedom or to the spins. Below we study the magnetic susceptibility, to explore which of these is responsible. We find that the spins are largely unaffected when we go through $T_{FL}$, thereby implicating the  charge degrees of freedom.

\subsection{Magnetic Susceptibility \label{susceptibility}}

The uniform magnetic susceptibility close to the Mott transition, $n
\gtrsim 0.75$, is one of the more difficult variables to compute
reliably by any technique, since it is highly enhanced by Stoner
factors $ \chi_{spin}/\chi^0_{spin}\sim 10$. In the ECFL theory we
found the numerical precision required for computing the
susceptibility hard to achieve with the scheme outlined in
\refdisp{ECFL-resistivity}. Although the local spectral functions for
either spin are confined to a compact region in frequency, it is their
difference that is needed for the susceptibility. This difference is
numerically very small and smeared over a large frequency range 
making it very difficult to control. The magnetic susceptibility
$\chi$ is a sensitive variable also within the DMFT using the NRG as
the impurity solver, in particular away from half filling at low
temperatures, thus it is seldom studied using this approach (see,
however \refdisp{Jarrell:1994ut,kajueter1996} for some very early DMFT
results, and \refdisp{bauer2009} for a more recent study using the DMFT (NRG) of
the half-filled Hubbard model in magnetic field at $T=0$). With some
effort we have found it possible to estimate its temperature
dependence. We used the method of finite field
\cite{DMFT,rozenberg1994} with $H=10^{-4}D \ll T$, which is
small enough for the system to remain well inside the linear response
regime, but sufficiently large to be little affected by numerical
noise. As a further test, we redid some calculations for $H=10^{-3}D$,
finding good consistency of the results.

\begin{figure}[h]
  \centering  
\includegraphics[width= 0.95\columnwidth]{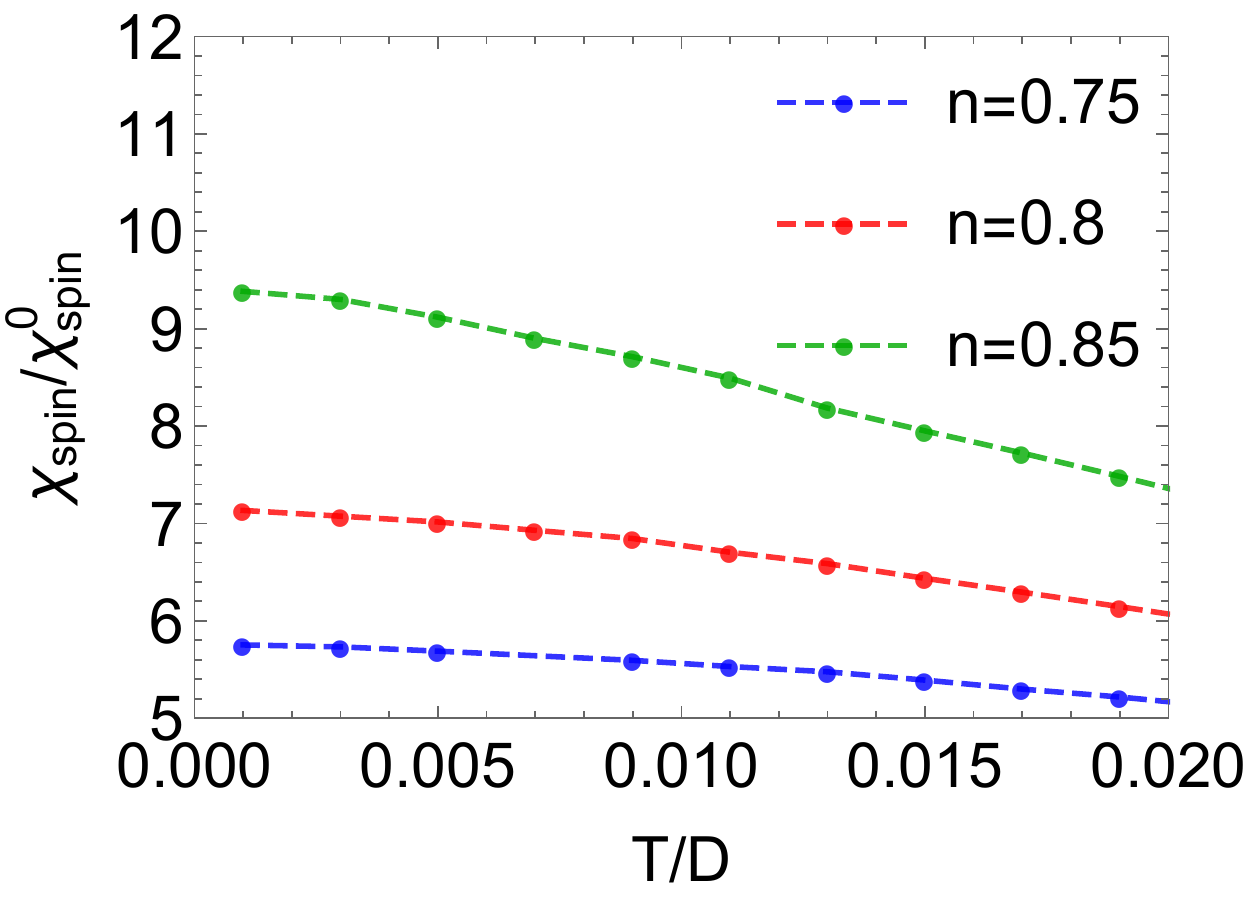}
 \caption{ Magnetic susceptibility (DMFT results). We note that the
 Stoner enhancement grows as $\delta\to 0$ and its $T$ dependence is
 Pauli like, but with a  somewhat enhanced $T$ dependence at higher $n$.
 The crossover to linear resistivity occurs (see \figdisp{fig:rho-compare}) at fairly low $T\lessim .005 D$ at these densities but has no reflection on the variation of $\chi_{spin}$. We may thus infer that spin disordering is not relevant to the linear resistivity seen here.  }
\label{fig:mag-chi-DMFT}
\end{figure}

In \figdisp{fig:mag-chi-DMFT} we present the DMFT Stoner enhancement of the susceptibility
$\chi_{spin}/\chi^0_{spin}$ as a function of $T$. Here the  spin susceptibility is denoted by $\chi_{spin}$ and for the noninteracting band case it is given by $\chi^0_{spin}= 2 \mu_B^2
{\cal D}(\varepsilon_F)$, where ${\cal D}$ is  the band density of states per spin per
site defined earlier.
 The scale of the Stoner enhancement is rather large, $\sim 10$. We find that the $T\to0$ value is roughly consistent with $1/Z$, as expected for an almost localized
Fermi liquid \cite{Vollhardt-RMP}. 

It is interesting that the Stoner factor and hence $\chi_{spin}$ is
Pauli-like in the temperature range studied here, i.e., the GCFL and
the GCSM regimes. It does not reflect the change in the resistivity
behavior from quadratic to linear. Thus the magnetic contribution to
the entropy change in \figdisp{fig:entropy-Cv} is very small, and we
must infer that the GCSM regime continues to have a quenched spin
entropy, as in the Fermi liquid. It would appear, by inference, that
the entropy released at $T_{FL}$ is charge related and the crossover
from the Fermi liquid to the GCSM may be viewed as partial charge
disordering. This is to be contrasted to the cross-over from GCSM to
the higher temperature bad metal regime, where the spin degrees of
freedom do become partially unscreened \cite{kokalj2013,dasari2016}.


\subsection{Thermoelectric transport \label{thermoelectric}}

\begin{figure}
  \centering
  \subfigure[]{\label{subfig:thermal-conductivity} \includegraphics[width=.8\columnwidth]{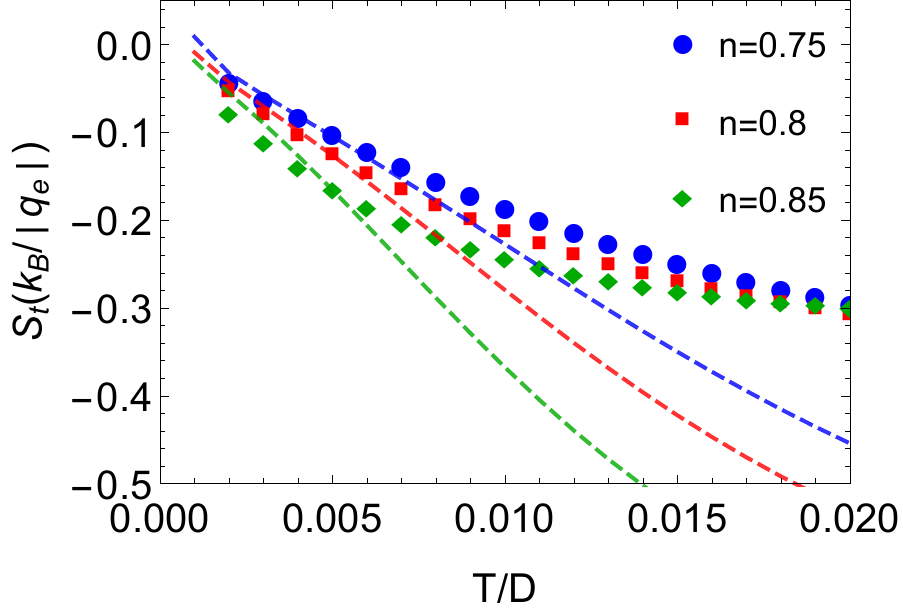}}
  \subfigure[]{\label{subfig:thermal-conductivity}  \includegraphics[width=.8\columnwidth]{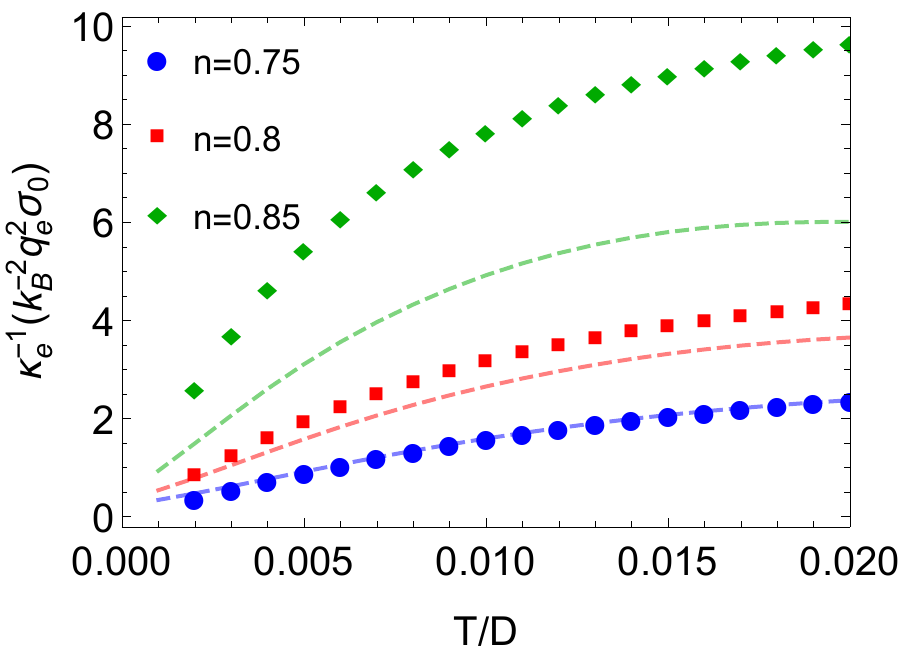}}
  \caption{ (a) Thermopower of ECFL (symbols) and DMFT (dashed lines).
  Both amplitudes and temperature derivatives are similar for  $T \leq .005$ but depart at higher T. (b) Electronic thermal resistivity $\kappa^{-1}_e$ of ECFL (symbols) and DMFT (dashed lines).}
  \label{fig:thermopower}
\end{figure}

\begin{figure}
  \centering
 \subfigure[]{  \includegraphics[width=.8\columnwidth]{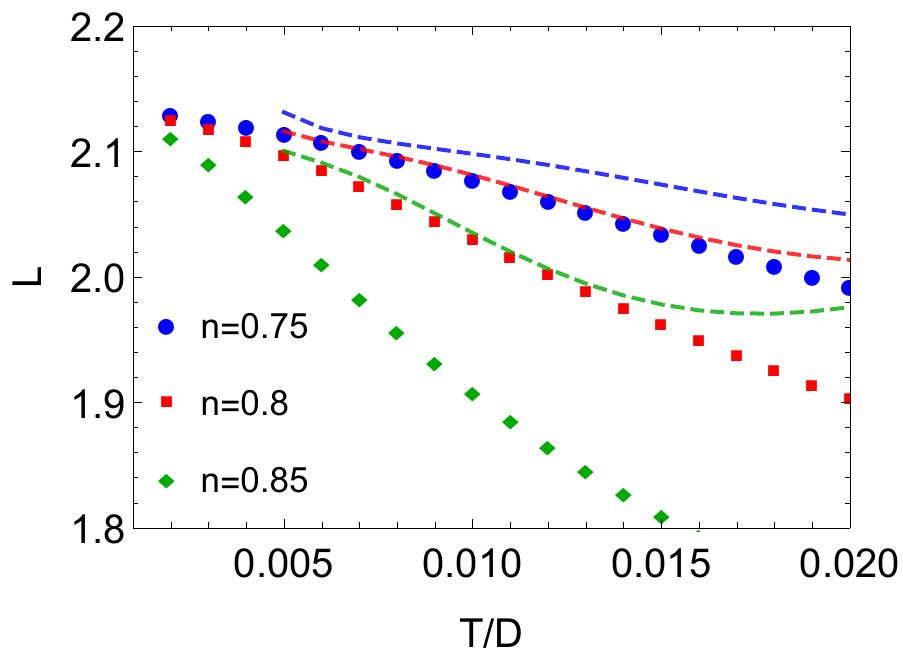}}
   \subfigure[]{\label{subfig:fog}\includegraphics[width=.8\columnwidth]{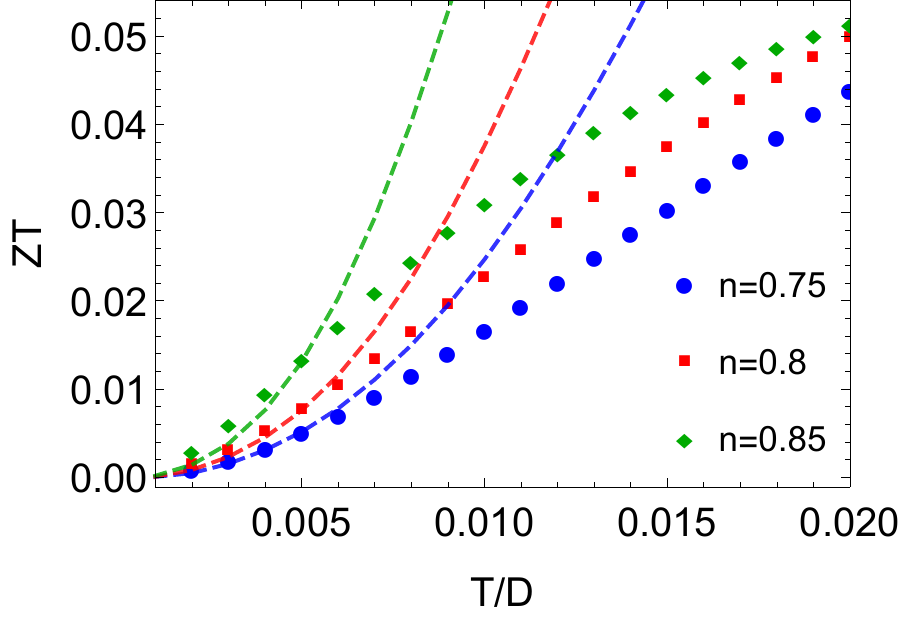}}
   \caption{(a) Lorenz number of  ECFL (symbols) and DMFT
   (dashed lines). The Lorenz number saturates to a constant ($\simeq 2.1$) which is typically expected for a Fermi liquid at low temperatures. (b) Figure of merit for ECFL (symbols) and DMFT (dashed lines). The low values of $ZT$ found here are typical of normal metals.}
   \label{fig:Lorenz-num}
\end{figure}

For completeness we present the results for the thermopower $S_t$, the
electronic thermal conductivity $\kappa_e$, and the Lorenz number $L$, as
well as the thermoelectric figure of merit in Fig. \ref{fig:thermopower} and \ref{fig:Lorenz-num}.  We record the expressions
following from standard transport theory\cite{Shastry-reports};
the thermopower $S_t$ and electronic thermal conductivity $\kappa_e$
are expressed in terms of three Onsager transport coefficients
$L_{11},\ L_{12} = L_{21}$, and $L_{22}$ as follows:
\begin{eqnarray}
  \sigma_{xx} = e^2 L_{11},\\
  S_{t} = - \frac{k_B}{\abs{e} T} \frac{L_{12}}{L_{11}},\\
  \kappa_e = \frac{k_B^2}{T}\Big(L_{22} - \frac{L_{12}^2}{L_{11}} \Big).
\end{eqnarray}
In infinite dimensions, these can be found in a straightforward way
from the spectral functions due to vanishing vertex corrections:
\begin{equation}
  L_{ij} = \frac{\sigma_0}{e^2} \int d\omega (- f'(\omega))
  \omega^{i+j-2} \int d\epsilon\, \Phi_{xx}(\epsilon) A^2(\omega,\epsilon).
\end{equation}
The Lorenz number is
\begin{equation}
  L=\frac{e^2}{k_B^2}\frac{\kappa_e}{\sigma_{xx} T},
\end{equation}
and the electronic thermoelectric figure of merit
\begin{equation}
  ZT = T\sigma_{xx} S_t^2 / \kappa_e
\end{equation}
(see Fig \ref{fig:thermopower}).

In the usual Fermi liquid theory, the electronic thermal conductivity
$\kappa_e \sim T^{-1}$ and the thermopower $S_t \sim \gamma T$. The
classic Lorenz number for a gas of particles with constant relaxation
time is $L_0 = \pi^2/3$ when we set $k_B = |e| = 1$, while for Fermi
liquid one expects $L_\mathrm{FL}=L_0/1.54 \approx 2.13$
\cite{herring}. In previous DMFT studies \cite{merino2000,
Shastry-reports, zlatic2012, zlatic2014}, thermal transport
coefficients were studied focusing on the very high temperature regime
of the bad metals. While our results qualitatively agree with the
previous studies, the crossover of thermal transport coefficients from
GCFL to GCSM in the low-$T$ regime (relative to the very high-$T$ bad
metal regime) are resolved. Both the thermopower
and thermal resistivity of ECFL change slope near $T_{FL}$. In DMFT
calculation, only the thermal resistivity shows similar crossover
behavior, while the thermopower seems to be insensitive to the
crossover from GCFL to GCSM. The Lorenz number of both ECFL and DMFT
converges to $L \simeq 2.1$ in the low-$T$ limit, as expected for a
Fermi-liquid ground state. The low values of $ZT$, shown in
Fig.~\ref{subfig:fog}, are typical of normal metals.

\section{Conclusions \label{sec-2}}

This work achieves two goals. On one hand, we explored the
low-temperature transport regimes of lattice fermions with the
constraint of no double occupancy (Gutzwiller projection) in the limit
of infinite dimensions.  We focus on the temperature range where the
Fermi-liquid quadratic resistivity gives way to the first $T$ linear
regime that we dubbed Gutzwiller correlated strange metal; this
cross-over occurs on the temperature scale which is much lower
compared with the bandwidth (and the Brinkman-Rice scale), but which
actually corresponds to the experimentally most relevant range of
order \unit[100]{K}. On the other hand, this work had a further
methodological goal of comparing the results for a number of
transport, spectroscopic and thermodynamic quantities obtained using
the mostly analytical extremely correlated Fermi liquid (ECFL) theory
and the accurate numerical results from the dynamical mean field
theory (DMFT) approach based on the numerical renormalization group as
the impurity solver. We found that at the cross-over temperature scale
both techniques indicate a change of behavior in most of the
quantities we investigated. The two methods have generally good
agreement, which improves upon lowering either the temperature or the
density.


The origin of the cross-over in the resistivity has been tracked down
to the temperature dependence of $-\Im m \Sigma(0,T)$, the imaginary part of the self-energy on the Fermi-surface, which starts to deviate from its low-temperature
asymptotic behavior on the scale $T_{FL}$ (Fermi-liquid temperature). This
low-energy scale is produced by purely local Gutzwiller correlation
effects, i.e., it is a direct consequence of the constraint of no
double occupancy of the lattice sites. We managed to show that $\rho(T)\propto - \Im m \Sigma(0,T)$ [\disp{ressimple}], which accounts well for the $\rho(T)$
dependence in the (GCFL)-Fermi-liquid and (GCSM)-strange metal regimes. As a result, we are able to explain the temperature dependence of the resistivity in terms of the temperature-dependence of the imaginary part of the self-energy on the Fermi surface.

The charge compressibility of the DMFT theory at infinite U is seen  to differ somewhat from that of the ECFL and also from the almost localized Fermi liquid. Developments in ECFL are underway in order to resolve the difference from DMFT.   The  compressibility shows a kink on the scale of $T_{FL}$ and
the heat capacity has a weak peak. The magnetic susceptibility,
however, shows no change across this cross-over. The cross-over hence
seems to be related to the charge degrees of freedom, while the spin
entropy is quenched in both Fermi liquid and strange metal regimes.
 It thus seems that the GCSM regime has a highly unusual composition, with some disordering of the charges, presumably in anticipation of the incipient Mott insulating state, without the participation of the spins.

{In a following paper, \refdisp{Hall-TBP},  we  present results for the
dynamical Hall constant and Hall angle indicating that the
two-relaxation-time behavior in transport properties observed in a
number of cuprates emerges upon entering the GCSM regime. Finally we note  a recent paper, \refdisp{Shastry-Mai}, where the results of a two-dimensional version of the equations studied here are presented.}


\begin{thebibliography}{99}
\bibitem{Gutzexplanation} {In the Gutzwiller approximation \cite{GA1,GA2}, double occupancy is projected out of the the Gutzwiller wave-function, while spin correlations are accounted for by purely combinatorial means. In Refs. \cite{MetznerVollhardt1,MetznerVollhardt2}, this approximation is derived diagrammatically, and shown to be exact in the limit of infinite spatial dimensions. In lower spatial dimensions, spin-correlations play a greater role, and hence the full Gutzwiller wave-function, with finite double-occupancy, must be used.}
\bibitem{chemical-shift} The chemical potentials  at the lowest $T$ for the three densities in \figdisp{fig:mu} are $0.72742, 0.98132, 1.35263$ for ECFL and $0.43986, 0.53668, 0.63789$ for DMFT.
\end{thebibliography}
\section{Acknowledgements} The work at UCSC was supported by the U.S.
Department of Energy (DOE), Office of Science, Basic Energy Sciences
(BES) under Award \# DE-FG02-06ER46319. R\v{Z} acknowledges the
financial support from the Slovenian Research Agency (research core
funding No.~P1-0044 and project No.~J1-7259). We thank Patrick Lee for
a stimulating discussion and Dieter Vollhardt for a helpful
correspondence. This work used the Extreme Science and Engineering Discovery Environment (XSEDE\cite{xsede} TG-DMR160144), which is supported by National Science Foundation Grant No. ACI-1053575, and the UCSC supercomputer Hyades, which is supported by National Science Foundation (award number AST-1229745) and UCSC.

\end{document}